\documentclass[aps,pra,twocolumn,superscriptaddress,longbibliography]{revtex4}
\usepackage{graphicx}
\usepackage{dcolumn}
\usepackage{bm}
\usepackage{physics}
\usepackage{amsmath}
\usepackage{amssymb}
\usepackage{array}
\usepackage{color}
\usepackage{times}
\newcolumntype{P}[1]{>{\centering\arraybackslash}p{#1}}
\usepackage[colorlinks=true, breaklinks=true, linkcolor=blue, citecolor=blue, urlcolor=blue]{hyperref}

\newcommand{\bonnpi}{Physikalisches Institut, University of Bonn, Nussallee 12, 53115 Bonn, Germany}
\newcommand{\geneva}{Department of Quantum Matter Physics, University of Geneva, Quai Ernest-Ansermet 24, 1211 Geneva, Switzerland}
\newcommand{\unbc}{Department of Physics, University of Northern British Columbia, Prince George, BC V2N 4Z9 Canada}
\newcommand{\dortmund}{Department of Physics, TU Dortmund University, 44227 Dortmund, Germany}
\newcommand{\koln}{Institute of Physics II, University of Cologne, 50937 Cologne, Germany}
\newcommand{\ausburg}{Experimental Physics V, Center for Electronic Correlations and Magnetism, Institute of Physics, University of Augsburg, 86135 Augsburg, Germany}

\begin{document}

\title{Repulsively bound magnon excitations of a spin-1/2 XXZ chain in a staggered transverse field}
\date{\today}

\begin{abstract}
  We study the excitation spectrum of the one-dimensional spin-1/2 XXZ chain with antiferromagnetic Ising anisotropy across a
  magnetic quantum phase transition induced by the application of a site-dependent transverse magnetic field.
  Motivated by the chain antiferromagnet BaCo\textsubscript{2}V\textsubscript{2}O\textsubscript{8}, we consider a
  situation where the transverse magnetic field has a strong uniform component and a weaker staggered part.
  To determine the nature of the excitations giving rise to the spin dynamical structure factor, we use a combination of
  analytical approaches and the numerically exact time-dependent matrix product state method. We identify below the quantum phase
  transition high-energy many-body two-magnon and three-magnon repulsively bound states which are
  clearly visible due to the staggered component of the magnetic field. At high magnetic fields and low temperature,
  single magnons dominate the dynamics. Our theory results are in very good agreement with terahertz spectroscopy experimental
  results presented in Ref.~\cite{WangKollath2024}.
\end{abstract}
\author{Catalin-Mihai Halati}
\affiliation{\geneva}
\affiliation{\bonnpi}
\author{Zhe Wang}
\affiliation{\dortmund}
\affiliation{\koln}
\affiliation{\ausburg}
\author{Thomas Lorenz}
\affiliation{\koln}
\author{Corinna Kollath}
\affiliation{\bonnpi}
\author{Jean-Sébastien Bernier}
\affiliation{\unbc}
\affiliation{\bonnpi}

\maketitle

\section{Introduction}

Quantum magnets are characterized by their delicate balance between multiple
competing interactions. Small modifications to this internal equilibrium often results
in significant changes to the system properties making these materials ideal
candidates to explore a vast realm of quantum phases, both conventional magnetically ordered ones
as well as unconventional phases exhibiting, for example, fractional excitations.
The range of possibilities grows even further as the quantum phases of several magnetic materials can be altered by the application
of external magnetic fields. While the required field strengths are not always
experimentally accessible, the discovery of low-dimensional quantum magnets
possessing relatively weak exchange interactions ~\cite{LakeFrost2013, MourigalRonnow2013, ZapfBatista2014, DallaPiazzaRonnow2015, ToskovicOtte2016, WuAronson2016, WangLoidl2016, BreunigLorenz2017, Wang2018, FaureGrenier2018, WangLorenz2019,Amelin_2022} has opened up novel avenues for exploration.

For example, these newly synthesized magnets were found to be particularly
suitable to investigate the physics of field-induced quantum phase transitions.
In recent years, a combination of experimental and theoretical studies considering such systems helped unveil the scaling properties at quantum critical points~\cite{BlosserZheludev2017, BlosserZheludev2018}, the presence of fractionalized excitations~\cite{ZaliznyakTakagi2004, ThielemannMesot2009b} and the occurrence of topological phase transitions~\cite{GiamarchiTchernyshyov2008}. These materials were also used to validate various aspects of Tomonaga-Luttinger liquid theory~\cite{KlanjsekGiamarchi2008, BouillotGiamarchi2011, SchmidigerZheludev2013}.

A compound of particular interest is the chain antiferromagnet BaCo\textsubscript{2}V\textsubscript{2}O\textsubscript{8}~\cite{HeItoh2005, KimuraItoh2008, CanevetLejay2013, KimuraWatanabe2013, NiesenLorenz2013, GrenierLejay2015, KlanjsekOrignac2015,FaureGrenier2018, WangLoidl2018b, FaureSimonet2021, WangKollath2024}. 
In this material, the Co$^{2+}$ ions are arranged in screw chains running along the fourfold $c$-axis of a body-centered tetragonal structure. 
While below the N\'eel critical temperature and at low applied magnetic fields, this material exhibits long-range antiferromagnetic ordering due to the presence of weak interchain couplings~\cite{NiesenLorenz2013,WangLoidl2018b}, at larger fields it effectively behaves, in many aspects,
as a quasi-one-dimensional system and can be modeled as a collection of weakly coupled spin-$1/2$ XXZ chains. 
When the external field is applied in the longitudinal direction, this compound presents a commensurate-incommensurate quantum phase transition between a N\'eel ordered phase and an incommensurate spin density wave phase \cite{KlanjsekOrignac2015, FaurePetit2019, TakayoshiGiamarchi2023}. Furthermore, for a specific window of longitudinal field strength and Ising anisotropy, high-energy many-body string excitations were experimentally detected~\cite{YangWu2019, WangLorenz2019}.

Due to the screw-chain structure of BaCo\textsubscript{2}V\textsubscript{2}O\textsubscript{8}, when an external field is applied in the transverse direction, the strength and even the direction of the effective local magnetic field can be staggered. 
While the study of the XXZ spin chain in a transverse magnetic field has been the subject of a number of works~\cite{DmitrievLangari2002, CauxLoew2003, BruognoloGarst2016, TakayoshiGiamarchi2018}, the effect of a staggered transverse field has received much less attention. 
The characteristics of the low-lying excitations were investigated in Ref.~\cite{OkutaniHagiwara2021} and the presence of a topological quantum phase transition described by a double sine-Gordon model when the local effective transverse field is fully staggered was reported in Ref.~\cite{TakayoshiGiamarchi2018}.

Here we consider the spin-$1/2$ XXZ chains with an Ising anisotropy subjected to a transverse field which contains both uniform and staggered components.
This is motivated by the application of an external field along the [110] direction in BaCo\textsubscript{2}V\textsubscript{2}O\textsubscript{8} which
leads to such an effective field \cite{KimuraWatanabe2013}.
For this case, a quantum phase transition belonging to the transverse-field Ising-chain universality class \cite{WangLoidl2018b} 
occurs between a phase with antiferromagnetic ordering in the direction perpendicular to the transverse field and a polarized state where
spins are aligned along the field direction. We are interested in understanding the nature of the excitations on both sides of this
transition for a wide range of energies. Through the computation and analysis of the dynamical spin structure factor, we identify
high-energy many-body two-magnon and three-magnon repulsively bound states below the quantum phase transition,
whereas above the transition we find that single-magnon excitations govern the quantum spin dynamics. 
We point out the importance of the staggered component of the magnetic field in clearly identifying the
repulsively bound states.
  In contrast to their attractively bound counterparts, repulsively bound states have higher energies than their unbound components.
  These interesting states have been observed, for example, in cold atomic gases \cite{WinklerZoller2006,DeuchertCederbaum2012}. 
  In closed quantum systems, repulsively bound states are long-lived as energy redistribution cannot easily be achieved. 
  It is interesting that in BaCo\textsubscript{2}V\textsubscript{2}O\textsubscript{8} such bound states, theoretically predicted within our work, can be observed experimentally \cite{WangKollath2024}, as in solid state systems the presence of multiple coupling channels could render these states much more unstable.
  
The article is structured as follows: in Sec.~\ref{sec:theomodel}, we introduce the theoretical model; in Sec.~\ref{sec:magnetization}, we characterize the phase transition taking place under the application of an external magnetic field when the effective local field has both uniform and staggered components;
in Sec.~\ref{sec:corr}, we introduce the relevant correlations; in Sec.~\ref{sec:numerical_results}, we present our numerical computation of the dynamical structure factor based on the time-dependent matrix product state method; 
in Sec.~\ref{sec:analytics}, we develop an analytical understanding of the excitations uncovered in the previous section; we finally conclude in Sec.~\ref{sec:conclusion}.

\section{Theoretical model \label{sec:theomodel}}

\begin{figure}[hbtp]
	\centering
	 \includegraphics[width=0.4\textwidth]{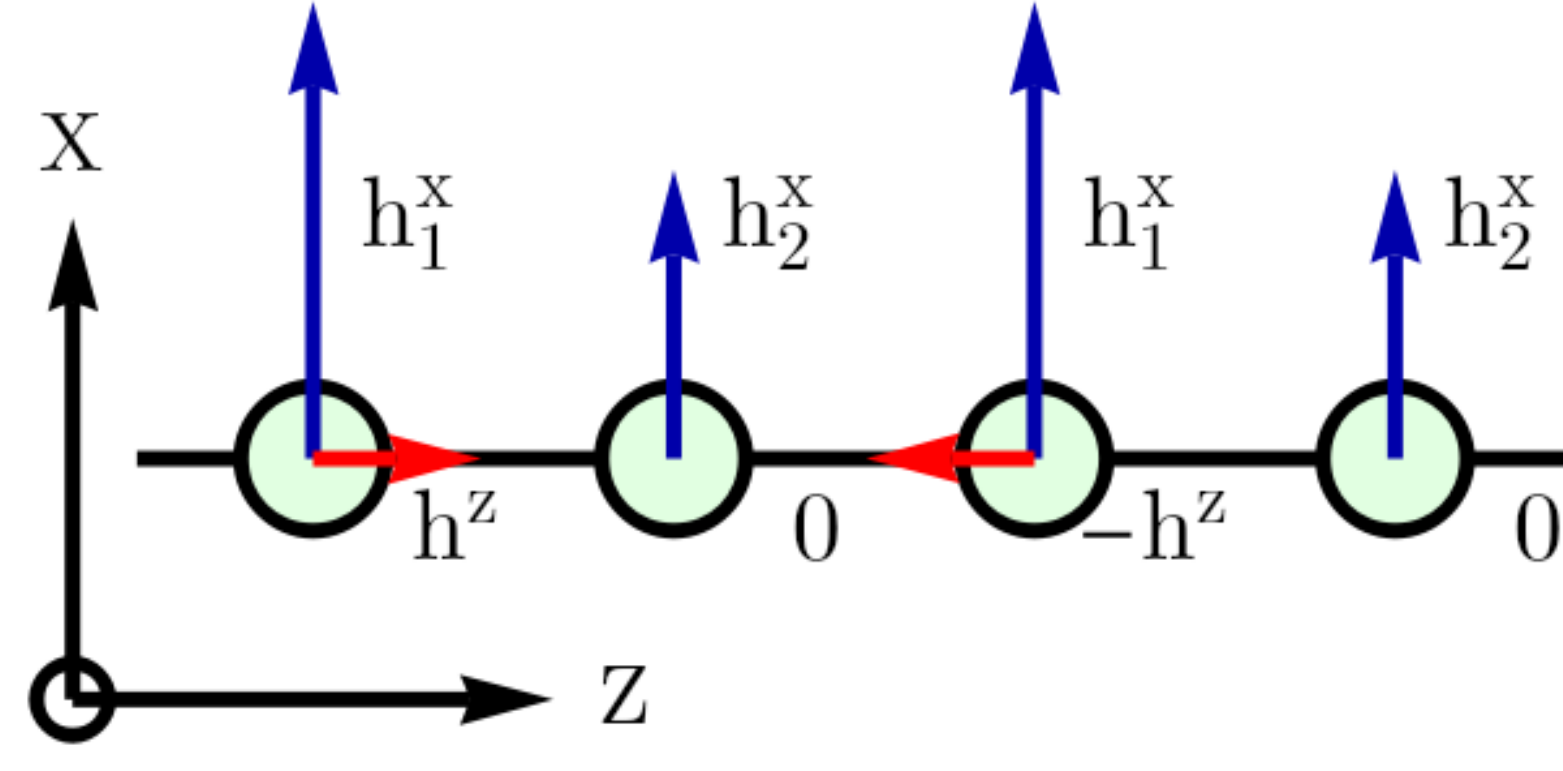} 
	\caption{Sketch of the space dependence of the effective magnetic field, Eqs.~(\ref{eq:Hamiltonian})-(\ref{eq:Hamiltonian2}). In the $x$-direction the magnetic field (blue arrows) has a uniform and a staggered component resulting in an alternating field amplitude from site to site. The field in the $z$-direction (red arrows) has a four-fold periodicity and takes finite values only on the odd sites.}
	\label{fig:field}
\end{figure}

In this work, we consider a one-dimensional spin $1/2$ chain subjected to a space dependent transverse magnetic field. 
This model is motivated by recent works on the
BaCo\textsubscript{2}V\textsubscript{2}O\textsubscript{8} compound \cite{HeItoh2005, KimuraItoh2008, CanevetLejay2013, KimuraWatanabe2013, NiesenLorenz2013, GrenierLejay2015, WangKollath2024}. BaCo\textsubscript{2}V\textsubscript{2}O\textsubscript{8}  belongs to a class of compounds with strong one-dimensional character exhibiting strong couplings within each spin chain and only very weak coupling between the chains.
Each chain is then modeled as a one-dimensional XXZ spin model with antiferromagnetic Ising anisotropy.
The resulting Hamiltonian for a single chain can be written in the form
\begin{align}
\label{eq:Hamiltonian}
H &=J\sum_j \left[ \epsilon\qty(S_j^x S^x_{j+1}+ S_j^y S^y_{j+1}) +S_j^z S^z_{j+1}\right] \\
&-h\sum_j\left(g^{x}_jS_j^x +  g^{z}_j S_j^z \right), \nonumber 
\end{align}
where we consider the following space dependence of the magnetic field
\begin{align}
\label{eq:Hamiltonian2}
g^x_j &= g^x_u-(-1)^j g^x_s, \\
g^z_j &=g^z\sin\left(\frac{\pi j}{2}\right). \nonumber
\end{align}
The magnetic field in the spin chain has a longitudinal and a transversal component (see sketch in Fig.~\ref{fig:field}). The transversal component,
along the $x$-direction, is chosen as the dominating field direction.
The transverse field component has a uniform part, as described by the factor $g_u^x$ on top of which a weaker part alternating in sign from site to site is added, Eq.~(\ref{eq:Hamiltonian2}). This weaker part is described by the site-dependent factor $(-1)^j~g^x_s$.
The longitudinal field along the $z$-direction will be typically chosen much smaller than the transversal part. It has a four-fold periodicity with non-zero and alternating values on the odd sites of the chain.

The site dependence of the magnetic field presented here models the application of an external magnetic field along the [110]
direction in the BaCo\textsubscript{2}V\textsubscript{2}O\textsubscript{8} compound. 
The form of the resulting effective field along the spin-$1/2$ chains in Eq.~(\ref{eq:Hamiltonian2}) is derived considering the Land\'e $g$-factors 
(see Ref.~\cite{KimuraWatanabe2013} and Appendix~\ref{app:gfactors}) considering a magnetic field applied along the $x$-direction and stems from the four-fold periodicity of the local Ising quantization axis in BaCo\textsubscript{2}V\textsubscript{2}O\textsubscript{8}.

In many realistic compounds, including BaCo\textsubscript{2}V\textsubscript{2}O\textsubscript{8}, an additional weak interchain coupling is present. Within a mean-field treatment, this interchain coupling corresponds to an additional magnetic field which needs to be self-consistently determined (see e.g.~\cite{MikeskaKolezhuk2004,Giamarchibook}). 
However, typically, its contribution is relatively small, being mainly of importance in the low external field regime
or  when considering low energy excitations at higher magnetic fields. 
We focus here on high energy excitations occuring in the large magnetic field regime near a phase transition towards the field-polarized state.
Consequently, we neglect the interchain coupling and concentrate on the investigation of the single chain Hamiltonian \cite{WangLoidl2018b,WangKollath2024}.

In the absence of a magnetic field, the one-dimensional XXZ chain shows a quantum phase transition at zero temperature with the anisotropy parameter $\epsilon$ \cite{MikeskaKolezhuk2004}. For small value of $|\epsilon|<1$, the system is in a ferromagnetic or antiferromagnetic gapped phase depending on the sign of the coupling $J$. For $|\epsilon|>1$, a gapless Luttinger liquid phase occurs. 
Here we set the value of the anisotropy parameter to $\epsilon=0.52$ and $J$ is a positive coupling.
The ground state in the absence of the magnetic field therefore lies in the antiferromagnetic phase.
 These parameter values were optimized in order to describe as best as possible experimental measurements of the
BaCo\textsubscript{2}V\textsubscript{2}O\textsubscript{8} compound~\cite{WangKollath2024}.

The application of an external magnetic field along the longitudinal or transverse direction of the XXZ model leads to a more complex phase diagram. 
For example, starting from the antiferromagnetic ground state at zero field and applying a longitudinal field,
a phase transition occurs to a Tomonaga-Luttinger liquid phase before the system reaches a fully polarized state at large magnetic fields \cite{MikeskaKolezhuk2004}.
For the case of a transverse uniform field, one has a transition from the antiferromagnet to a  fully polarized state in the direction of the transverse field~\cite{Sachdevbook,DmitrievLangari2002}.
In contrast, if the transverse field is fully staggered, i.e. changes sign from site to site, a topological transition occurs~\cite{MikeskaKolezhuk2004,Sachdevbook, TakayoshiGiamarchi2018}.
The transitions for both a uniform, or a staggered, transverse field are in the Ising universality class~\cite{Sachdevbook, DmitrievLangari2002, TakayoshiGiamarchi2018}.
In this work, we consider a transverse field that contains both uniform and staggered components. As we consider a dominant uniform component,
we observe a phase transition with a similar behavior as for the case of a purely uniform magnetic field.

In order to obtain unbiased results for the model given in Eqs.~(\ref{eq:Hamiltonian})-(\ref{eq:Hamiltonian2}), we employ standard matrix product state techniques \cite{Schollwoeck2011}. These are variational methods which approximates the targeted state by a so-called matrix product state (MPS) with a maximal matrix dimension. This approximation corresponds to using states with a low von-Neumann entropy. Matrix product state methods have been shown to be extremely successful for the description of low dimensional spin chains both for equilibrium and dynamic properties. 
The numerical results for the ground state of the model, e.g.~for the magnetization (Sec.~\ref{sec:magnetization}), were obtained using a finite-size density matrix renormalization group (DMRG) algorithm in the matrix product state (MPS) representation \cite{White1992,Schollwoeck2005,Schollwoeck2011, Hallberg2006,Jeckelmann2002}, implemented using the ITensor Library \cite{FishmanStoudenmire2020}. The convergence is ensured by a maximal bond dimension up to 300, for which the truncation error is at most $10^{-12}$.

Performing calculations using the MPS algorithms with different values for the parameters $g^x_u$, $g^x_s$, $g^z$, the anisotropy $\epsilon$, and the spin coupling $J$, we identified the set of parameters for which the dynamic structure factor and the magnetization curves agree best with the experimental measurements of the BaCo\textsubscript{2}V\textsubscript{2}O\textsubscript{8} compound \cite{WangKollath2024} and are consistent with previous experimental studies \cite{KimuraWatanabe2013, FaureGrenier2018,WangLorenz2019}.
Most of the results we present in the following are obtained using  $g^x_u=3.06$, $g^x_s=0.66$, $g^z=0.21$ and $\epsilon=0.52$, if not stated otherwise. However, for a theoretical understanding of the influence of the different parameters, we also present results where we vary these parameters. 
We note that these parameters correspond to the values $g_1=3.7$, $g_2 =6.1$, $g_3 =2.4$, $\theta = 5^{\circ}$, $\Delta=1.92$ in the notation
of Ref.~\cite{WangKollath2024}.
 In order to ease comparison between the results presented here and the ones of Ref.~\cite{WangKollath2024}, for a coupling $J=5.43~\text{meV}$, the conversion goes as follow: a frequency of $\hbar\omega = J$ corresponds to $\omega=1.31~\text{THz}$ and a magnetic field of $h =0.1 J$ corresponds to $B=9.38~\text{T}$.

\section{Magnetization behavior \label{sec:magnetization}}

\begin{figure}[hbtp]
\centering
\includegraphics[width=0.45\textwidth]{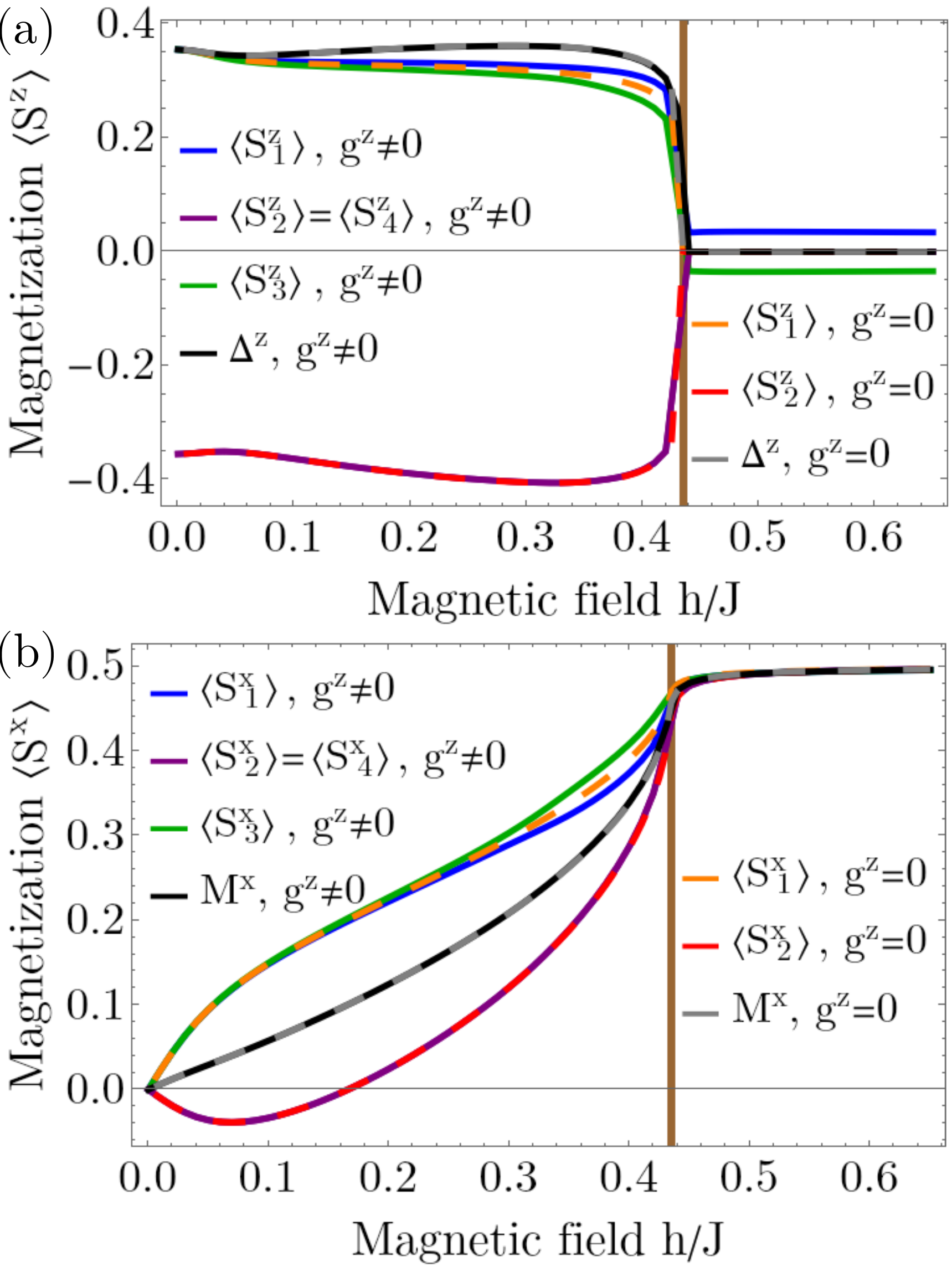} 
\caption{The magnetization of the individual sites of the unit cell, averaged over the entire system length, in the ground state of Eqs.~(\ref{eq:Hamiltonian})-(\ref{eq:Hamiltonian2}), for (a) $z$-direction and (b) $x$-direction.
In panel (a) we also compute the staggered magnetization $\Delta^z$ and in panel (b) the average magnetization $M^x$, defined in Eq.~(\ref{eq:mag}).
The numerical DMRG results where obtained for the following parameters, $L=124$, $g^x_u=3.06$, $g^x_s=0.66$, $g^z=0.21$ and $\epsilon=0.52$.
The dashed lines correspond to the case of $g^z=0$. The vertical line at $h_c/J=0.435$ marks the transition threshold.}
\label{fig:magnetization}
\end{figure}

For the model considered in this work, Eqs.~(\ref{eq:Hamiltonian})-(\ref{eq:Hamiltonian2}), a phase transition occurs, with increasing the magnetic field,
between an antiferromagnetic state in the $z$-direction and a fully polarized state in the $x$-direction, the direction of the applied magnetic field~\cite{KimuraWatanabe2013}.
This can be observed by computing the average and staggered magnetizations in these two directions \cite{KimuraWatanabe2013}
\begin{align}
\label{eq:mag}
M^x&=\frac{1}{L}\sum_j \langle S^x_j \rangle, \\
\Delta^z&=\frac{1}{L}\sum_j (-1)^j \langle S^z_j \rangle. \nonumber 
\end{align}

For the parameters considered here, we observe in Fig.~\ref{fig:magnetization} that as one increases the magnetic field, the ground state of the system remains in an antiferromagnetic state in the $z$-direction up to $h_c/J\lesssim 0.435$. 
This can be inferred from the large value of the staggered magnetization $\Delta^z$ shown in Fig.~\ref{fig:magnetization}(a) up to $h_c/J\approx 0.435$ and the vanishing value for larger magnetic fields. 
Considering the four local  magnetizations $\langle S^z_j \rangle$, we observe, on even sites, the same
  negative value below the transition while  these two values vanish above the threshold.
However, for the odd sites, due to the presence of a small magnetic field in the $z$-direction, the local magnetization on the
first and third site of the unit cell are slightly different whereas, above the transition threshold, they take small values with opposite signs.
The magnetization in the field direction, $M^x$, is increasing with the magnetic field and has a large value close to the phase transition, see Fig.~\ref{fig:magnetization}(b).
Above the phase transition $h_c/J\approx 0.435$ the system becomes almost fully polarized in the field direction and the magnetization $M^x$ saturates to a value close to $0.5$.
Due to the strong staggered ($g_s^x$) contribution in the magnetic field term and the strength of the spin-spin interaction along the field direction (see Sec.~\ref{sec:noninteracting}), at small field values, $h/J\lesssim 0.15$, the local magnetizations $\langle S^x_j \rangle$ have opposite signs on the odd and even sites, signaling a tendency towards antiferromagnetic ordering in the $x$-direction. However, this is suppressed by the uniform component of the magnetic field in the $x$-direction, such that already for $h/J\gtrsim 0.2$ the magnetization $\langle S^x_j \rangle\gtrsim 0.25$ for odd $j$.
For the even sites, the local magnetization starts to increase rapidly as one gets closer to the critical field.
The difference between the local magnetizations on the odd and even sites would be less prominent for smaller values of the staggering strength $g_s^x$.
Similarly to the $z$-direction, near the phase transition, the different values of $\langle S^x_j \rangle$ on the first and third
sites of the unit cell are due to the four-fold field $h g^z_j$.

The position of the phase transition threshold is controlled mostly by the interplay between the strength of the uniform and staggered
components of the magnetic field, $g_u^x$ and $g_s^x$, and the anisotropy $\epsilon$ (see Sec.~\ref{sec:1mag}).
We observe that by neglecting the small value of magnetic field in the $z$-direction [dashed lines in Fig.~\ref{fig:magnetization}(a)-(b)],
 $M^x$ and $\Delta^z$
remain almost the same while the local magnetizations on the two odd sites adopt the same value (this occurs for both spin directions).

\section{Dynamical Observables of interest \label{sec:corr}}

In order to capture the properties of the spin excitations across the quantum phase transition, we compute
the dynamical spin structure factor.  Besides the instrinsic theoretical interest, this quantity
  can be directly related using linear response theory to experimental results obtained via terahertz spectroscopy technique,
  as we show in Ref.~\cite{WangKollath2024}, or other experimental techniques such as neutron scattering.
The dynamical structure factors for the different spin directions are defined as
\begin{align}
\label{eq:corr1}
S^{\alpha \beta}_l(q,\omega) =\frac{1}{\sqrt{L}}\int_0^\infty \sum_{j=1}^L e^{i(\omega t-qj)}S^{\alpha \beta}_{j,l}(t) dt,
\end{align}
with $q= \frac{2\pi k}{L}$, $k=0\dots L-1$ the discrete quasi-momenta in the first Brillouin zone, and $\alpha$ and $\beta$ the two spin directions.

The quasi-momentum $q$ has the reciprocal units of the lattice spacing $a$, which we set to $a=1$. 
 For the system considered here, Eqs.~(\ref{eq:Hamiltonian})-(\ref{eq:Hamiltonian2}), the unit cell for the lattice
  with the applied magnetic field contains four sites.
However, when the weak effective field contribution along the $z$-direction is neglected (see Sec.~\ref{sec:numerical_results}), the unit cell only consists of two lattice sites and at zero magnetic field the unit cell is just a single lattice site.
This would lead to different first Brillouin zones corresponding to the different unit cells. To avoid changing the
  representation, the results are represented by unfolding the band structure to the so-called extended Brillouin zone.
  The quasi-momentum is chosen from $0$ to $2\pi$, in the following, we loosely refer to quasi-momentum $q$ as momentum.
We use here the exponential Fourier transformation for infinite systems, since we apply a Gaussian filter in order to minimize the effects of the open boundary conditions (see below).

In order to determine the spin structure factor from MPS simulations, we use the time-dependent MPS method based on Trotter-Suzuki decompositions \cite{DaleyVidal2004,WhiteFeiguin2004}.
As it is needed in Eq.~(\ref{eq:corr1}), the first step consists in computing the two-point spin correlation functions at different moments in time
\begin{align}
\label{eq:corr2}
S^{\alpha \beta}_{j,l}(t)=\bra{0} S^\alpha_j (t) S^\beta_l\ket{0}=\bra{0} e^{i t H/\hbar} S^\alpha_j e^{-i t H/\hbar} S^\beta_l\ket{0},
\end{align}
with $\ket{0}$ the ground state of the spin chain, unless specified otherwise. 
We perform the time-evolution of the excited states $\ket{\psi_l}=S^\beta_l\ket{0}$ and compute the overlap of $S^\alpha_j\ket{\psi_l (t)}$
and $e^{-i t E_0/\hbar}\ket{0}$, with $E_0$ the ground state energy. The time evolution of the ground state is not explicitly performed, but the energy $E_0$ is
taken from a ground state calculation which reduces the numerical cost of the computation.
The sites for which we compute the correlations are $l\in\{L/2-1,L/2,L/2+1,L/2+2\}$ corresponding to the central four-site unit cell and taking the site $j$ over the entire chain, i.e.~$j=1\dots L$. 

From these time-dependent and position-dependent quantities, we obtain the dynamical structure factors via a numerical Fourier transform to the frequency-momentum space
\begin{align}
\label{eq:bacovo_corr}
S^{\alpha \beta}_l(q,\omega) =\frac{1}{\sqrt{N_t L}}\sum_{n=0}^{N_t-1} \sum_{j=1}^L e^{i(\omega n \delta t-qj)} f(j) S^{\alpha \beta}_{j,l}(n \delta t),
\end{align}
with the discrete momenta $q= \frac{2\pi k}{L}$, $k=0\dots L-1$ and frequencies $\omega=\frac{2\pi s}{N_t \delta t}$, with $N_t$ the number of time measurements and $\delta t$ the time interval between them, $s=0\dots N_t-1$. 
We reduce the effects of the open boundary conditions by applying a Gaussian filter to the dynamic correlations before the numerical Fourier transform
\begin{align}
\label{eq:corr_filter}
f(j)= e^{-4\left( 1-\frac{2j}{L-1} \right)^ 2}.
\end{align}
The filter minimizes the numerical artifacts arising due to the use of open boundary conditions, but, at the same time, this filtering
reduces the momentum resolution. The width of the Gaussian filter was chosen to balance out these two effects.

  In order to make the interpretation of the momentum and frequency resolved dynamical structure
  factor $S^{\alpha \beta}_l(q,\omega)$, Eq.~(\ref{eq:bacovo_corr}), clearer, we note that its
  spectral representation is 
\begin{align}
\label{eq:spectral_representation}
S^{\alpha \beta}_l(q,\omega) =\sum_{e} \bra{0} S^\alpha (q) \ket{e}&\bra{e} S^\beta_l \ket{0}  \\
&\times\delta(\omega+E_0-E_e), \nonumber
\end{align}
where $\ket{e}$ are the eigenstates of the considered Hamiltonian and $E_e$ the corresponding eigenenergies.
The Fourier transform of the spin operator is defined as $S^\alpha(q)=\frac{1}{\sqrt{L}}\sum_{j=1}^L e^{-iqj} f(j)S^\alpha_j$.
Thus, the dynamical observable $S^{\alpha \beta}_l(q,\omega)$ connects the eigenstates of the Hamiltonian with the
ground state via the application of the operators $S^\beta_l$ and $S^\alpha(q)$.
In other words, the spin structure factor measures the amplitude of making
an excitation at a well defined quasi-momentum $q$ and energy $\omega$ starting from the ground state via the
spin operator $S^\beta$. The $\delta$-function ensures energy conservation.

In Sec.~\ref{sec:numerical_results}, we present the numerical results in which we averaged over $l$ in one unit cell
\begin{align}
\label{eq:corr3}
\mathcal{S}^{\alpha \beta}(q,\omega) =\sum_{l=L/2-1}^{L/2+2} \left|S^{\alpha \beta}_l(q,\omega)\right|^2.
\end{align}
This is the quantity that we will call the dynamical structure factor for the rest of this study
and we focus on the $\alpha=\beta=z$ component which displays rich features.
We checked that for the other combinations perpendicular
to the $x$-direction of the magnetic field, $(\alpha,\beta)\in\{(y,y),(y,z),(z,y)\}$,
no strong new features arise (see Sec.~\ref{sec:numerical_results}).
In fact, the spin structure factor for $\alpha=\beta=z$ contains the important features seen in terahertz spectroscopy measurements. In Ref.~\cite{WangKollath2024}, these features were shown to been related to a combination of dynamical correlation functions in the plane transverse to the external magnetic field.

We compute numerically the dynamical correlations using the time-dependent matrix product state method
(tMPS)~\cite{DaleyVidal2004, WhiteFeiguin2004, Schollwoeck2011}. For the details of our implementation see Ref.~\cite{HalatiPhD}.
To carry out the simulation of time-dependent quantities, we typically consider systems made of $L = 124$ sites and bond dimensions up to 300 states.
This ensures that the truncation error at the final time $t J/\hbar=110$ is $\lesssim 10^{-7}$. We note that at large magnetic
fields, above the phase transition, the truncation error at the final time is even smaller, $\lesssim 10^{-10}$.
The convergence was ensured with a time step of $\mathrm{d}t J/\hbar=0.05$ 
and the measurements were performed every fourth time step.
Unless stated otherwise, we make use of open boundary conditions for the spin chain. 

\begin{figure}[t]
\centering
\includegraphics[width=0.48\textwidth]{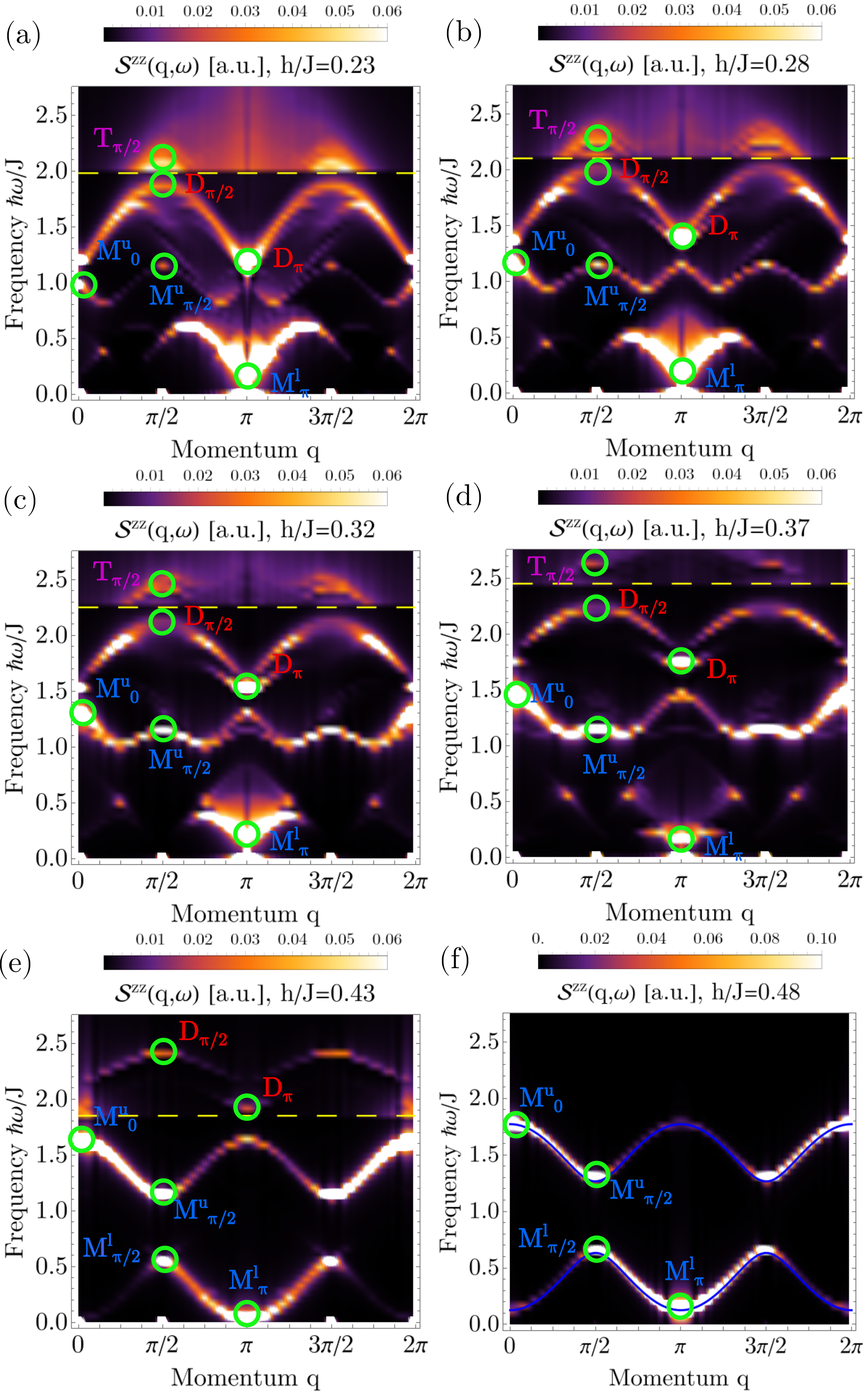} 
\caption{The dynamical structure factor for different values of the magnetic field as a function of momentum $q$ and frequency $\omega$ for the model given in Eq.~(\ref{eq:Hamiltonian}).
In each panel we normalized the maximum of $\mathcal{S}^{zz}(q,\omega)$ to $1$.
Above the horizontal yellow dashed line we multiplied $\mathcal{S}^{zz}(q,\omega)$ with a factor of 10 in order to increase the visibility of the high-frequency modes.
We use the green circles to mark the modes we focus our analysis. In panel (f) the blue curves are given by the one-magnon dispersion relation, Eq.~(\ref{eq:bands_one_magnon})
The numerical tMPS results where obtained for the following parameters, $L=124$, $g^x_u=3.06$, $g^x_s=0.66$, $g^z=0.21$ and $\epsilon=0.52$.
The values of the magnetic field presented correspond to $B\in\{21.6,26.3,30,34.7,40.3,45\}~\text{T}$.}
\label{fig:spectra}
\end{figure}

\section{Numerical \lowercase{t}MPS results - Momentum dependent spectral functions\label{sec:numerical_results}}

\begin{figure}[t]
\centering
\includegraphics[width=0.48\textwidth]{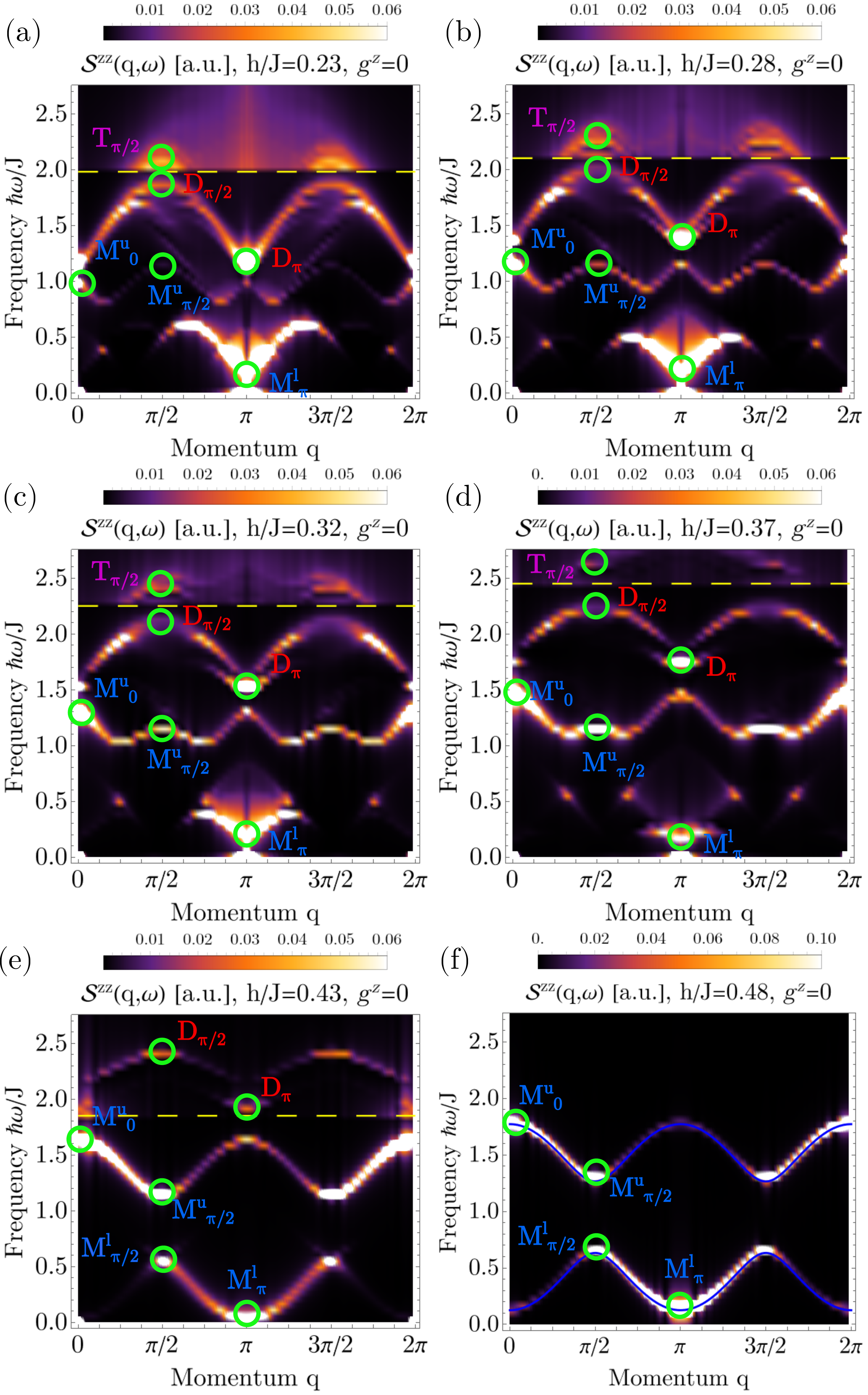} 
\caption{The dynamical structure factor for different values of the magnetic field as a function of momentum $q$ and frequency $\omega$, in the case of neglecting the field in the $z$-direction, $g^z=0$.
In each panel we normalized the maximum of $\mathcal{S}^{zz}(q,\omega)$ to $1$.
Above the horizontal yellow dashed line we multiplied $\mathcal{S}^{zz}(q,\omega)$ with a factor of 10 in order to increase the visibility of the high-frequency modes.
We use the green circles to mark the modes we focus our analysis. In panel (f) the blue curves are given by the one-magnon dispersion relation, Eq.~(\ref{eq:bands_one_magnon})
The numerical tMPS results where obtained for the following parameters, $L=124$, $g^x_u=3.06$, $g^x_s=0.66$, $g^z=0$ and $\epsilon=0.52$.
The values of the magnetic field presented correspond to $B\in\{21.6,26.3,30,34.7,40.3,45\}~\text{T}$.}
\label{fig:spectra_gz0}
\end{figure}

\begin{figure*}[!ht]
\centering
\includegraphics[width=0.93\textwidth]{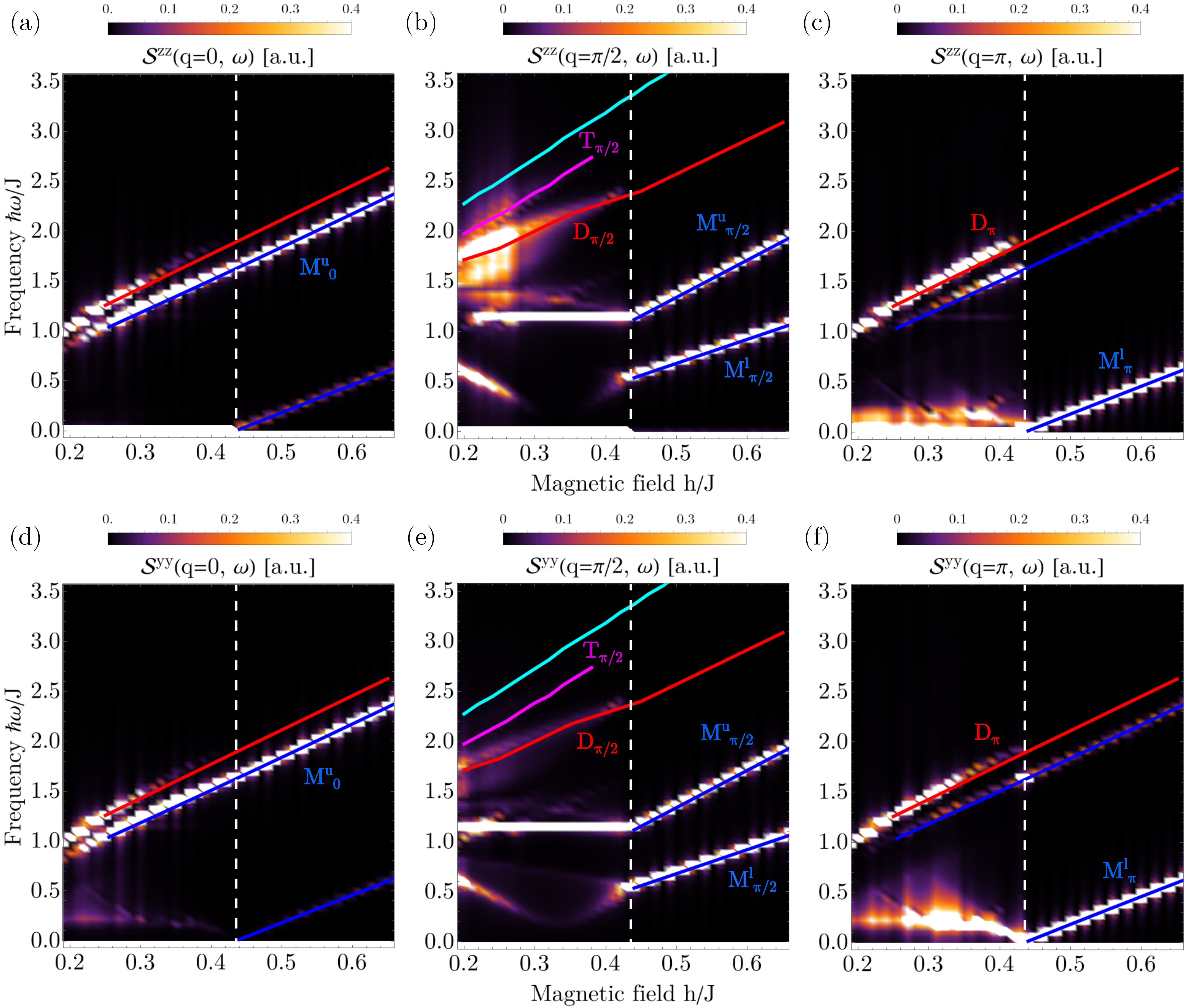} 
\caption[Summary plot for the dynamical structure factor]{
Comparison between the numerical computation of the dynamical structure factor (a)-(c) $\mathcal{S}^{zz}(q,\omega)$, (d)-(f) $\mathcal{S}^{yy}(q,\omega)$, Eq.~(\ref{eq:corr3}), for the full Hamiltonian, Eq.~(\ref{eq:Hamiltonian}),  the one-magnon analytical dispersion, Eq.~(\ref{eq:bands_one_magnon}), (blue lines), the $q=\pi/2$ and $q=\pi$ peaks of the two-magnon states contribution to the dynamical structure factor (red lines - see Sec.~\ref{sec:2mag}), and  $q=\pi/2$ peaks of the three-magnon states contribution to the dynamical structure factor (cyan and magenta lines - see Sec.~\ref{sec:3mag}). 
The vertical white line represents the value of the critical magnetic field at $h/J\approx 0.435$.
We normalize $\mathcal{S}^{\alpha\alpha}(q,\omega)$ for each value of the magnetic field $h/J$, such that the maximum as a function of the frequency $\hbar\omega/J$ is set to $1$.
The numerical tMPS results where obtained for the following parameters, $L=124$, $g^x_u=3.06$, $g^x_s=0.66$, $g^z=0.21$ and $\epsilon=0.52$.
}
\label{fig:summary}
\end{figure*}

In this section, we present the numerical tMPS results for the dynamical spin structure
factor $\mathcal{S}^{zz}(q,\omega)$. We obtain its full dependence on the momentum $q$ and frequency $\omega$, as presented
in Fig.~\ref{fig:spectra} for several values of the magnetic field.

As a first step, we analyze the influence of the small $z$ component of the magnetic field, Eq.~(\ref{eq:Hamiltonian}). We use in Fig.~\ref{fig:spectra} a value of $g^z=0.21$ in order to compare with the results for the BaCo\textsubscript{2}V\textsubscript{2}O\textsubscript{8} compound \cite{WangKollath2024}. In Fig.~\ref{fig:spectra_gz0} we neglect this term in the Hamiltonian by setting $g^z=0$. We observe that the main features of the spectral function remain the same. Only small differences occur such as minor redistributions of spectral weight for some features.
Therefore, we will be able to safely neglect this term when using analytical approaches to identify
  the nature of the main features present in the spectral function in Sec.~\ref{sec:analytics}.

In this analytical analysis,
we focus mainly on the features which could be observed in the terahertz spectroscopy
experiment presented in Ref.~\cite{WangKollath2024}. These are the modes which have an important weight for the values of quasi-momenta $q\in\{0,\pi/2,\pi\}$ in the extended Brillioun zone as these correspond to zero-momentum transfer. We mark these features in Fig.~\ref{fig:spectra} and depict their dependence with respect to the magnetic field in Fig.~\ref{fig:summary}. We observe that the minimum of the lowest band, marked by $M_\pi^l$, decreases as the transition is approached from below and the gap closes
at the critical value of the magnetic field, $h_c/J \approx 0.435$, signaling the phase transition [see Fig.~\ref{fig:spectra}(e) and Fig.~\ref{fig:summary}(c)]. The band shifts to higher frequencies as we increase the magnetic field beyond the transition.
Furthermore, in the lowest band a second local minimum exists between $q = 0$ and $q = \pi$, whose position changes as a function of the
magnetic field. The change in position is reflected in the mode we can observe in Fig.~\ref{fig:summary}(e) at $q=\pi/2$, which decreases in frequency until it reaches its smallest gap at around $h/J\approx 0.35$. 
At this value of the magnetic field, the minimum of the lowest band does occur at $q=\pi/2$.
Afterwards its frequency increases across the phase transition, marked with $M_{\pi/2}^l$ above $h_c$.

The second band, marked by $M_0^u$ and $M_{\pi/2}^u$, situated around $\hbar\omega/J\approx 1.1$ in Fig.~\ref{fig:spectra}(c), changes quite drastically in between $h/J=0.23$ and $h/J=0.43$, due to its change of curvature around $q=\pi/2$.
 While the peak at $q=\pi/2$ remains constant in frequency, Fig.~\ref{fig:summary}(b), the peaks at $q=0,\pi$ are increasing with the magnetic field.
Above the critical field, this band monotonously increases with the magnetic field to higher energies.
We see in Sec.~\ref{sec:analytics} that the lowest two bands above the critical field stem from the dynamics of single magnons in the staggered potential due to the  site dependence of the magnetic field.

The third band, which we can observe between $\hbar\omega/J\approx1.5$ and $\hbar\omega/J\approx2$ in Fig.~\ref{fig:spectra}(c) at $h/J=0.28$, increases in energy with the magnetic field, however close to the critical field its intensity is small compared to the low-frequency excitations. This band contains the modes labeled by $D_{\pi/2}$ and $D_\pi$, whose dependence on the magnetic field is shown in Fig.~\ref{fig:summary}(b)-(c). We determine its nature as being due to the dynamics of two-magnon bound states, which are repulsively bound (Sec.~\ref{sec:analytics}).
Furthermore, at even higher frequencies for $q=\pi/2$ we can observe a distinct feature, labeled by $T_{\pi/2}$, Fig.~\ref{fig:spectra}(a)-(d) and Fig.~\ref{fig:summary}(b).
 The frequency of the $T_{\pi/2}$ mode increases with the magnetic field and it is visible in our numerical results for fields up to $h/J\approx 0.4$. We identify this feature as appearing due to bound states of three magnons, see Sec.~\ref{sec:analytics}.

In Fig.~\ref{fig:summary}, we compare the dynamical structure factors $\mathcal{S}^{zz}(q,\omega)$ [Fig.~\ref{fig:summary}(a)-(c)] and $\mathcal{S}^{yy}(q,\omega)$ [Fig.~\ref{fig:summary}(d)-(f)] for the selected values of momenta $q\in\{0,\pi/2,\pi\}$ as a function of the magnetic field.
We observe that the main features we have identified in this section are present for both spin direction, only the weights of the peaks are different. In $\mathcal{S}^{yy}$, we
resolve better the low-energy modes, in particular the mode at $q=\pi$ which shows the closing of the gap as we approach the transition threshold from below.
However, the high energy modes $D_{\pi/2}$ and $T_{\pi/2}$ are more prominent in $\mathcal{S}^{zz}$.

Thus, we have observed in this section that the considered model, Eq.~(\ref{eq:Hamiltonian}), has a complex excitation spectrum (see Fig.~\ref{fig:spectra}). In order to gain a better understanding of the nature of these excitations, we employ a series of analytical approaches in the next section.

\section{Understanding the nature of the excitations via analytical approaches \label{sec:analytics}}

In this section, we consider the spin chain Hamiltonian $H$ in various limits to develop an analytical
understanding of the nature of the excitations present near the critical
magnetic field. We first neglect the term coupling the spins to an
induced magnetic field along the $z$-direction. This approximation is motivated by BaCo$_2$V$_2$O$_8$
as for the parameters used to describe this material the site-dependent $g$-factors along the $z$-direction are
$g^{z}_j=(0.21,0,-0.21,0)$, one order of magnitude smaller than the ones
along the $x$-direction, $g^{x}_j=(3.72,2.4,3.72,2.4)$.
Furthermore, we observed in Sec.~\ref{sec:numerical_results} that neglecting the field in the $z$-direction has
very small impact on the features present in the dynamical structure factor on which we focus in this work.

Within this first approximation, the Hamiltonian becomes
\begin{align}
\label{eq:bacovo_Hamiltonian_reduced}
\tilde{H} =&~J\sum_j \left[ S_j^z S^z_{j+1}+\epsilon\left(S_j^x S^x_{j+1}+ S_j^y S^y_{j+1}\right) \right] \\
&-h\sum_j \left[ g_u -(-1)^{j} g_s \right] S_j^x, \nonumber
\end{align}
where we dropped the $x$ label on the $g$-factors, $g_u\equiv g_u^x$ and $g_s\equiv g_s^x$.

In order to obtain insights into the behavior at moderate to high values of the magnetic fields, we
then perform a rotation aligning the direction of the external magnetic field with the new $z$-direction, i.e.
\begin{align}
\label{eq:bacovo_rot2}
S^x &\to S^z, \\
S^y &\to S^x,  \nonumber \\
S^z &\to S^y.  \nonumber 
\end{align}
This rotated Hamiltonian reads
\begin{align}
\label{eq:Hamiltonian_rotated}
\tilde{H} =&~J\sum_j \left[ S_j^y S^y_{j+1}+\epsilon\left(S_j^z S^z_{j+1}+ S_j^x S^x_{j+1}\right) \right] \\
&-h\sum_j \left[ g_u -(-1)^{j} g_s \right] S_j^z, \nonumber
\end{align}
and can be rewritten as
\begin{align}
\label{eq:bacovo_Hamiltonian_reduced_mag}
\tilde{H} =J\sum_j& \left[\epsilon S_j^z S^z_{j+1} +\frac{1}{4} (1+\epsilon) (S^+_j S^-_{j+1} + S^-_{j} S^+_{j+1}) \right. \nonumber \\
&- \left. \frac{1}{4}(1-\epsilon) (S^+_j S^+_{j+1} + S^-_{j} S^-_{j+1})\right] \nonumber \\
-h\sum_j& \left[ g_u -(-1)^{j} g_s \right] S_j^z,
\end{align}
using $S^x_j=\frac{1}{2}\left(S^+_j+S^-_j\right)$ and $S^y_j=\frac{1}{2i}\left(S^+_j-S^-_j\right)$.
The analysis presented in the rest of this section will make reference to this rotated Hamiltonian, $\tilde{H}$.
However, for the sake of clarity and to ease comparison with the numerical results, we will identify
in the various figures presented below the dynamical structure factor using the original spin directions.
The corresponding rotated directions will be provided in all captions to avoid any ambiguity.

The Hamiltonian $\tilde{H}$ of Eq.~(\ref{eq:bacovo_Hamiltonian_reduced_mag}) still describes a complicated
many-body system characterized by the the interplay between the space dependent magnetic field $\left[ g_u -(-1)^{j} g_s \right] S_j^z$,
the spin-spin interactions $S_j^z S^z_{j+1}$, the kinetic energy $(S^+_j S^-_{j+1} + S^-_{j} S^+_{j+1})$ and terms changing
the total magnetization $(S^+_j S^+_{j+1} + S^-_{j} S^-_{j+1})$. 
Therefore, in order to understand the excitation spectrum of this model, we need to consider several approximate
limits. In Sec.~\ref{sec:noninteracting}, we consider the non-interacting version of the Hamiltonian $\tilde{H}$, related to the Kitaev chain model,
which can be exactly solved, and show that the dynamics around the phase transition is dominated by states having a
small number of flipped spins. It should be noted that neglecting the interaction in the rotated basis does not amount
to neglecting the $zz$-coupling in the original basis. Then, in Sec.~\ref{sec:interacting}, we analyze the effects of the interactions by
considering different fixed magnetizations sectors of the Hamiltonian and reveal the presence of repulsively bound states made of
two or three confined magnons. We show that the insight gained from performing these approximations and the
resulting quantitative descriptions agree very well with the numerically exact calculations performed
on the full model, given in Eq.~(\ref{eq:Hamiltonian}). Thus, we are confident that these various approaches
capture the true physical nature of a category of excitation modes present in this system.

\subsection{Magnon excitations within the non-interacting model \label{sec:noninteracting}}

\begin{figure}[hbtp]
	\centering
	 \includegraphics[width=0.47\textwidth]{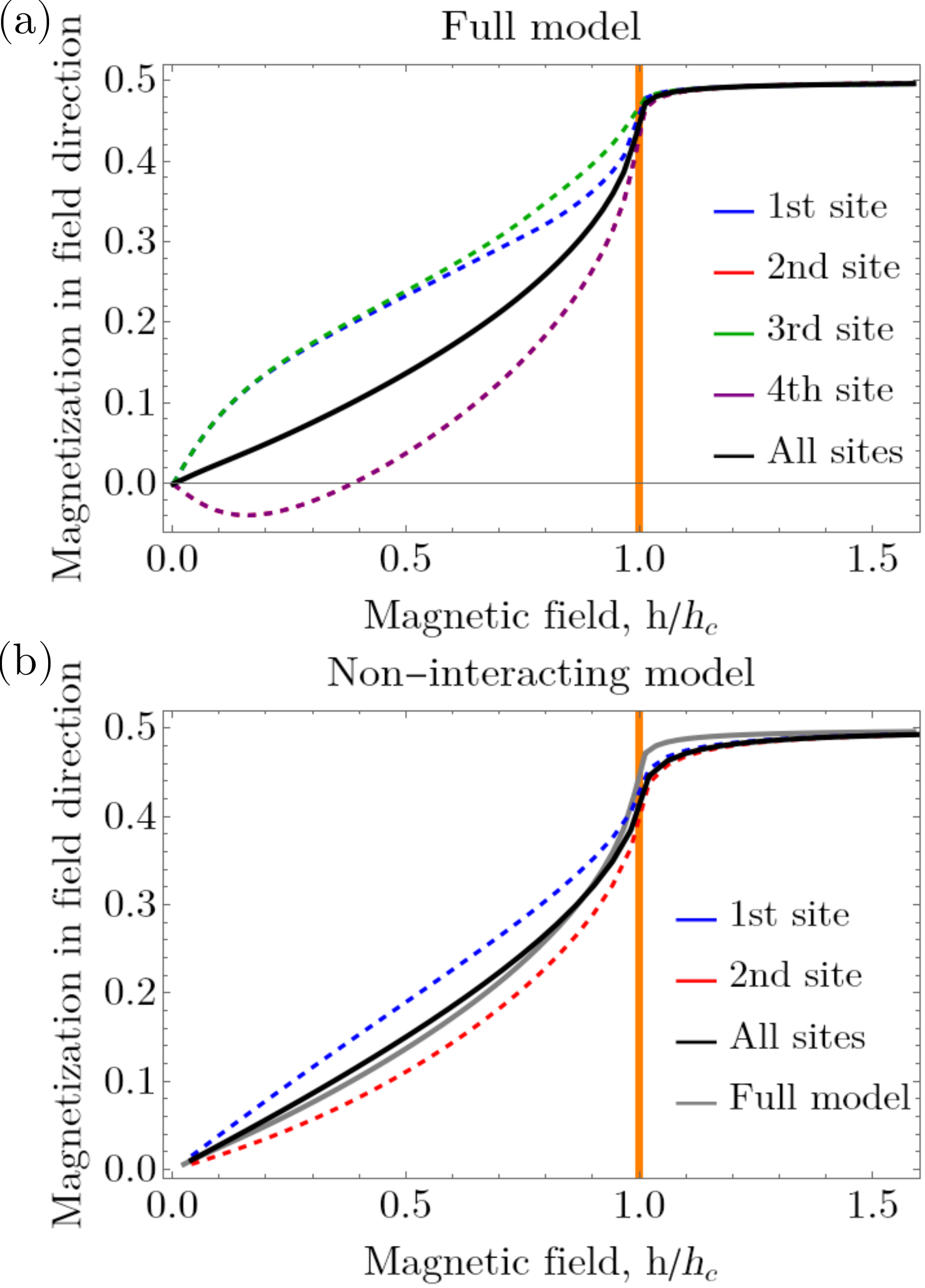} 
	\caption{The magnetization in the field direction averaged over the unit cell, or for the individual sites of the unit cell, for (a) the full model in Eq.~(\ref{eq:Hamiltonian}), and (b), the non-interacting model, Eq.~(\ref{eq:Hamiltonian_noninteracting}). 
The field direction corresponds to $x$-direction in the Hamiltonian given in Eq.~(\ref{eq:Hamiltonian}) and to the $z$-direction in the rotated frame Eqs.~(\ref{eq:bacovo_rot2})(\ref{eq:Hamiltonian_rotated}).
Black curve in panel (a) is repeated as the gray curve in panel (b) to be easier to compare the two regimes.
The results in (a) were obtained using DMRG ground state simulations and in (b) via analytical exact solution of the non-interacting model. The orange vertical line marks the critical magnetic field, which correspond to $h_c/J=0.435$ for the full model and $h_c/J=0.255$ for the non-interacting model. 
The parameters used are $L=124$, $g^x_u=3.06$, $g^x_s=0.66$, $g^z=0.21$ and $\epsilon=0.52$.  The curves in (a) for the 2nd site and the 4th site overlap with each other.}
	\label{fig:magnetization_noninteracting}
\end{figure}

We first investigate the excitations present in the non-interacting limit where the term corresponding to the spin-spin
interaction in the direction of the field $S_j^z S^z_{j+1}$ is neglected. Note, that this does not correspond to neglecting the interaction in the original model.
The resulting simple model is of interest
because it exhibits a very similar phase transition between an antiferromagnetic phase along the $y$-direction
and the field-polarized state also found when considering
$\tilde{H}$, Eq.~(\ref{eq:bacovo_Hamiltonian_reduced_mag}) (as discussed in Sec.~\ref{sec:magnetization} for the full model),
even though the two transitions occur at different critical magnetic field values due to the absence of the interaction term.

This non-interacting Hamiltonian reads
\begin{align}
\label{eq:Hamiltonian_noninteracting}
\tilde{H}_0 =~\frac{J}{4}\sum_j& \left[(1+\epsilon) (S^+_j S^-_{j+1} + S^-_{j} S^+_{j+1}) \right. \\
-& \left. (1-\epsilon) (S^+_j S^+_{j+1} + S^-_{j} S^-_{j+1})\right] \nonumber \\
-h\sum_j& \left[ g_u -(-1)^{j} g_s \right] S_j^z \nonumber.
\end{align}
$\tilde{H}_0$ can be solved exactly using the Jordan-Wigner transformation
\begin{align}
 \label{eq:JWtransformation}
  S^z_j &= \frac{1}{2} - c^\dagger_{j} c_j, \\
  S^+_j &= c_j e^{-i\pi\sum_{l=1}^{j-1} c^\dagger_{l} c_l}, \nonumber \\
  S^{-}_j &= e^{i\pi\sum_{l=1}^{j-1} c^\dagger_{l} c_l} c^\dagger_{j}, \nonumber
\end{align}
where $c^\dagger_{j}$ and $c_j$ are fermionic operators satisfying the anticommutation relations
$\{c_l, c^\dagger_j\} = \delta_{lj}$ and $\{c^\dagger_l, c^\dagger_j\} =  \{c_l, c_j\} = 0$.
Using this rewriting, a magnon at site $j$, defined as a flipped spin compared to the direction of the applied field, corresponds to the presence of a fermion,
and the non-interacting Hamiltonian takes the form
\begin{align}
\label{eq:Hamiltonian_noninteracting_fermion}
\tilde{H}_0 =~\frac{J}{4} \sum_j& \left[(1+\epsilon)\left(c^\dagger_j c_{j+1} + c^\dagger_{j+1} c_j \right) \right. \\
-& \left. (1-\epsilon) \left( c^\dagger_j c^\dagger_{j+1} + c_{j+1} c_j \right)  \right] \nonumber \\
- h\sum_j& \left[ g_u -(-1)^{j} g_s \right] \left(c^\dagger_j c_{j} - \frac{1}{2} \right). \nonumber
\end{align}
This formulation is particularly useful as it allows us to gain a deeper understanding of the nature
of the excitations on both sides of the phase transition. 
This model is related to the Kitaev chain for $g_s=0$ which shows a transition from a topological non-trivial phase to a trivial phase with increasing magnetic field \cite{GreiterThomale2014}.

\begin{figure}[hbtp]
	\centering
	 \includegraphics[width=0.47\textwidth]{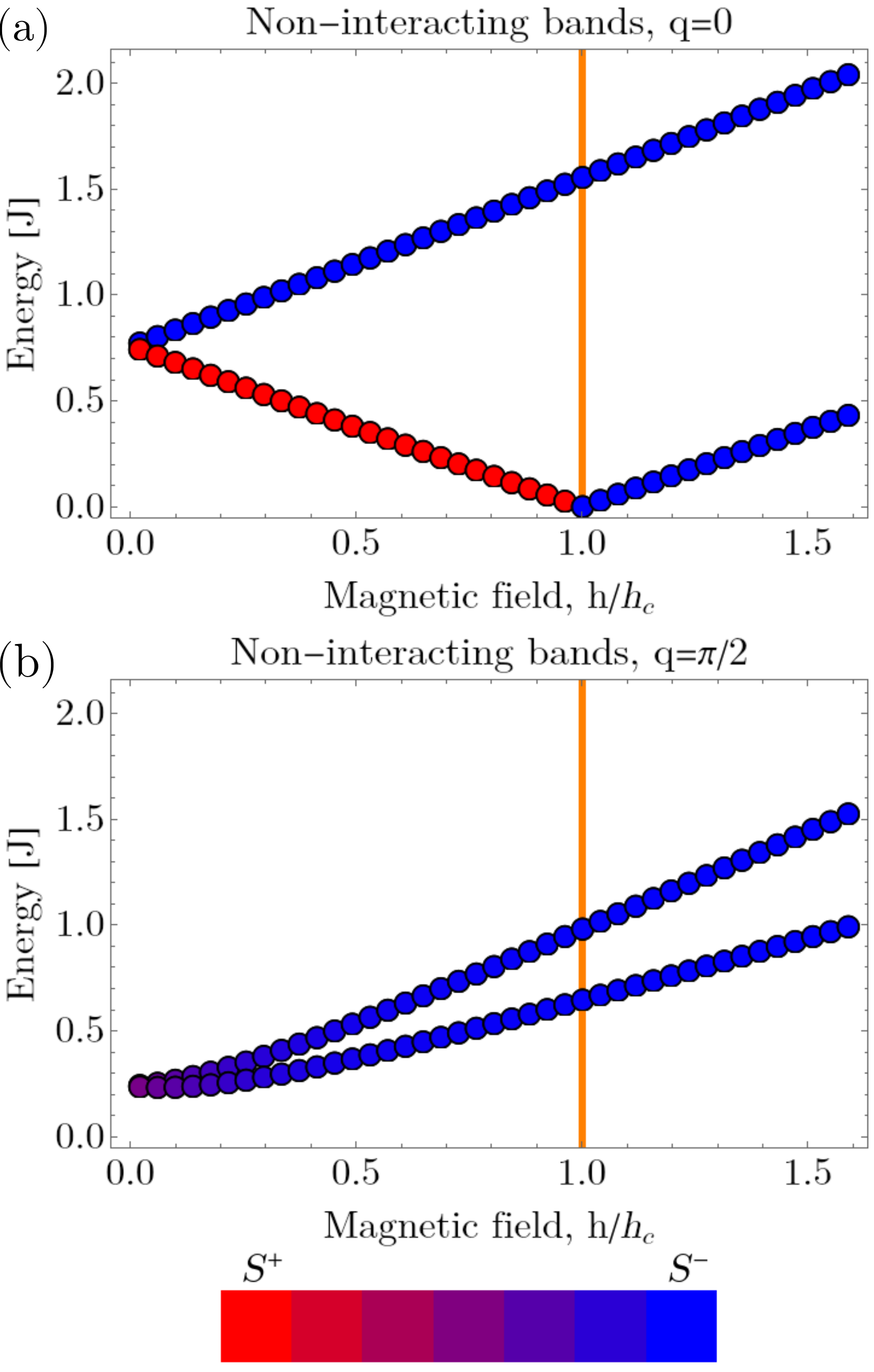} 
	\caption{The excitation bands of the non-interacting model for (a), $q=0$ and (b), $q=\pi/2$, obtained from the analytical exact solution of  Eq.~(\ref{eq:Hamiltonian_noninteracting}). Blue indicates that most of the weight of the excitation comes from $S^-$ and red that most of the weight comes from $S^+$, while purple corresponds to equal weight being shared between the two operators. The orange vertical line marks the critical magnetic field. 
Here $S^+$ and $S^-$ are considered in the rotated basis Eq.~(\ref{eq:bacovo_rot2}).
The parameters used are $L=124$, $g^x_u=3.06$, $g^x_s=0.66$ and $\epsilon=0.52$.}
	\label{fig:excitation_nature}
\end{figure}

It is instructive to first compare the magnetization along the field direction for the full Hamiltonian $\tilde{H}$,
Eq.~(\ref{eq:Hamiltonian}), and the non-interacting one
$\tilde{H}_0$, Eq.~(\ref{eq:Hamiltonian_noninteracting}). Since the value of the critical field is distinct for the two cases,
the field is shown scaled by the corresponding critical field $h_c$. As can be seen on Fig.~\ref{fig:magnetization_noninteracting}(a)
and Fig.~\ref{fig:magnetization_noninteracting}(b),
both magnetization curves behave very similarly. Above the critical field, both systems become almost fully magnetized along the field direction.
More importantly, the magnetization along the field direction decreases slowly below the critical field.
Thus, below the field-induced phase transition, there exists a large field range in which the magnetization along the
field direction remains substantial, even though these systems order in the perpendicular direction. Note that whereas the overall
magnetization behaves very similarly, the magnetization on the single sites behave slightly differently in the two situations.
In particular, the alternating behavior between even and odd sites is more enhanced by the presence of the interaction
in the full model $\tilde{H}$, Eq.~(\ref{eq:Hamiltonian}).

In both situations, above the critical field, the magnetization in the field direction
is almost maximal, such that the ground state for both $H$ and $\tilde{H}_0$ is very well approximated
by the fully polarized state. In this regime the excitations can be understood in terms of magnons, spin flips
in the polarized background or in the fermionic picture as a single fermion on the vacuum.
This interpretation is confirmed by considering the structure of the eigenmodes obtained by diagonalizing $\tilde{H}_0$, Eqs.~(\ref{eq:Hamiltonian_noninteracting})-(\ref{eq:Hamiltonian_noninteracting_fermion}).

In Fig.~\ref{fig:excitation_nature}, we show the excitation bands of $\tilde{H}_0$ across the phase transition for
$q = 0, \pi/2$ together with the weight of the $S^+$ and $S^-$ operators in the eigenmodes.
By diagonalizing the Hamiltonian given in Eq.~(\ref{eq:Hamiltonian_noninteracting_fermion}), one obtains
the eigenmodes as a combination of $c_j$ and $c_j^\dagger$ operators, i.e. $\gamma^\dagger_q=\sum_j \left( \alpha_{j,q} c_j + \beta_{j,q} c_j^\dagger \right)$,
with complex coefficients $\alpha_{j,q},\beta_{j,q}$ and normalization $\sum_j \left( |\alpha_{j,q}|^2+ |\beta_{j,q}|^2\right)=1$.
Thus, we can define the weight associated with all $c_j$ operators for a certain $q$, corresponding to the $S_j^+$ operators in the spin language, as the sum $\sum_j |\alpha_{j,q}|^2$,
and the weight associated to all $c_j^\dagger$ operators, corresponding to the $S_j^-$ operators, as the sum $\sum_j |\beta_{j,q}|^2$.
The following color coding is employed in Fig.~\ref{fig:excitation_nature}: Blue indicates that most of the weight comes from $S^-$
and red that of the most weight comes from $S^+$, while purple corresponds to
equal weight being shared between the operators. 
The action of the $S^-$ operator on the polarized state creates a magnon. Thus, we associate the weight of the $S^-$ operators with the ``magnon'' nature of the excitations. This interpretation is valid in the regime of high magnetization. 
The two distinct bands in Fig.~\ref{fig:excitation_nature} are due to the staggering in the magnetic field. 
We observe that above the critical field the excitations are of magnon character as almost all weight is due to $S^-$ operators.

When describing states with a large magnetization, one can efficiently use this description, as only a reduced number of magnons are relevant. In our case, we observe that the magnetization of the ground state is sizable
from $h\gtrsim 0.75 h_c$ corresponding to a dilute density of magnons 
both for the full and non-interacting models
[see Fig.~\ref{fig:magnetization_noninteracting}(a)(b)]. 
Whereas the character of the lowest excitations changes at the phase transition, the excitations of the upper band maintain their magnon character
even in the regime of large magnetic fields below the transition where $0.5h_c\lesssim h\lesssim h_c$.
In fact, it is only at noticeably lower magnetic fields [see Fig.~\ref{fig:excitation_nature}(b)] that the upper
eigenmodes are described by a combination of $S^+$ and $S^-$.

Thus, the analysis performed above shows that it is possible to describe the high-energy excitations below the
critical field and the excitations above the transition in terms of states with a small and well-defined number
of magnons. In the following section, we analyze the effects of the magnon-magnon interaction.

\subsection{Presence of magnon-magnon interactions \label{sec:interacting}}

In the magnetic field range $h > 0.5 h_c$ the ground state and excitations both above and below the critical field can be described in
terms of states having a magnon character, as we saw in Sec.~\ref{sec:noninteracting}. In the following, we analyze the effect
of the magnon-magnon interaction considering the Hamiltonian
\begin{align}
\label{eq:Hamiltonian_magnons}
\tilde{H}_\text{eff}=J &\sum_j \left[  \epsilon S_j^z S^z_{j+1} +\frac{1}{4}(\epsilon+1)\qty(S_j^+ S^-_{j+1}+ S_j^- S^+_{j+1}) \right] \nonumber \\
 -h&\sum_j \left[ g_u -(-1)^{j} g_s \right]  S_j^z.
\end{align}
When compared to $\tilde{H}$, we neglected here the term $(S_j^+ S^+_{j+1}+ S_j^- S^-_{j+1})$. This allows us to restrict
our analysis to sectors with a well-defined magnetization in the magnetic field direction. As we will show below, our analytical
results based on Eq.~(\ref{eq:Hamiltonian_magnons}) reproduce correctly the location and the field dependence
of the features present in the different spectra obtained when considering the full Hamiltonian via tMPS
(see Fig.~\ref{fig:summary}). This good agreement gives us confidence that our approximation is justified
for obtaining the nature of the modes. While the neglected term would connect these sectors, its presence would not alter
the nature of the observed modes, it would only affect the relative height of the peaks of the dynamical structure factor.
We further confirmed this by numerically computing the dynamical structure factor of $\tilde{H}$, Eq.~(\ref{eq:bacovo_Hamiltonian_reduced_mag}),
for which we reduced the strength of the $(S_j^+ S^+_{j+1}+ S_j^- S^-_{j+1})$ term by one, or two, orders of magnitude (not shown).
In these cases we observe that the position of the high-frequency modes is not changed, but only their amplitude is modified with the main impact being on the low-frequency modes under the critical field. 
Furthermore, the expectation value of the neglected terms in the ground state of the full model, Eq.~(\ref{eq:Hamiltonian}),
is always one order of magnitude smaller than the dominant contribution to the total energy, 
either the kinetic energy or the magnetic field term.

In the high field regime, the main mechanism for probing experimentally the excitations stems from flipping a spin,
creating a magnon \cite{WangKollath2024}. This experimental process motivates us to consider contributions to the structure factor for
transitions between sectors for which the number of magnons differs by one. In the following subsections, we
separately discuss the contributions stemming from the one-, two-, and three-magnon sectors.
We use the Jordan-Wigner transformatiom given in Eq.~(\ref{eq:JWtransformation}) to map $\tilde{H}_\text{eff}$ to
the following fermionic Hamiltonian
\begin{align}
\label{eq:Hamiltonian_magnons_fermions}
\tilde{H}_\text{eff} = J\sum_j& \Bigg[  \epsilon \qty(\frac{1}{2}-c^\dagger_j c_j)\qty(\frac{1}{2}-c^\dagger_{j+1} c_{j+1})  \\
&+ \frac{1}{4}(\epsilon+1)\qty(c^\dagger_j c_{j+1}+ c^\dagger_{j+1}c_j ) \Bigg] \nonumber\\
-h\sum_j & \left[ g_u -(-1)^{j} g_s \right]  \qty(\frac{1}{2}-c^\dagger_j c_j), \nonumber
\end{align}
where each magnon corresponds to the presence of a fermion at the same site. We use this representation to obtain
the dispersion relation in the one-magnon sector and to explore other multi-magnon sectors.

\subsubsection{One-magnon sector}
\label{sec:1mag}

We first consider the situation where a single magnon excitation is created out of the the totally polarized state.
This corresponds to a creation of a single fermion within the Hamiltonian presentation given in
Eq.~(\ref{eq:Hamiltonian_magnons_fermions}). In this case, we look for the signature of this excitation
within the dynamical spin structure factor by considering transitions from the ground state to
the one-magnon sector such that
\begin{align}
\label{eq:bacovo_corrdef}
&S^{\alpha,\beta}_{j}(q,\omega) \\
&~~\propto \sum_{e_{1m}} \bra{0}  S^\alpha(q) \ket{e_{1m}} \bra{e_{1m}} S^\beta_j\ket{0} \delta(\hbar\omega+E_{0}-E_{1m}), \nonumber
\end{align}
where  $\ket{0}$ is the fully polarized ground state with energy $E_{0}$, $\ket{e_{1m}}$ are the one-magnon eigenstates with the corresponding
eigenenergies $E_{1m}$, and $ S^\alpha(q)=\frac{1}{\sqrt{L}}\sum_l e^{-iql}S^\alpha_l$. Within the rotated basis, Eq.~(\ref{eq:bacovo_rot2}),
transitions between these sectors only occur for $\alpha,\beta\in\{x,y\}$.
We observe that $S^{\alpha,\beta}_{j}(q,\omega)$ is non-zero only when $\hbar\omega+E_{0}-E_{1m}=0$, thus, since $E_{0}$ is a constant,
we can capture the structure of $S^{\alpha,\beta}_{j}(q,\omega)$ by computing the dispersion
relation $E_{1m}(q)$ of one magnon (fermion) under the action of the fermionic Hamiltonian from Eq.~(\ref{eq:Hamiltonian_magnons_fermions}). 

The ground state energy of the Hamiltonian $\tilde{H}_\text{eff}$, corresponding to the fully polarized state (the fermionic vaccum state), is
\begin{align}
\label{eq:bacovo_egs}
E_{0} &= L\frac{\epsilon}{4}-L\frac{hg_u}{2}.
\end{align}
The dispersion in the one-magnon sector is (see eg.~\cite{Giamarchilecturenotes})
\begin{align}
\label{eq:bands_one_magnon}
E_{1m}^\pm(q) &= g_uh-J\epsilon\pm \sqrt{\varepsilon^0(q)^2+(g_sh)^2} \\
&= g_u h- J \epsilon \nonumber\\
&\pm \sqrt{\qty[\frac{J}{2}(1+\epsilon)\cos(q)]^2+(g_sh)^2} \nonumber,
\end{align}
where, in the last line, we replaced the kinetic dispersion by its explicit expression
$\varepsilon^0(q)=\frac{J}{2}(1+\epsilon)\cos(q)$. Whereas in a uniform magnetic field a pure cosine contribution
is expected in the one-magnon sector, the staggered contribution $g_s$ to the magnetic field splits the one-magnon
dispersion into two distinct bands, as shown below in Fig.~\ref{fig:spectra}(f) (solid line). 

The lower one-magnon band exhibits minima at $q=0$ and $q=\pi$ and their corresponding gap closes as we approach
the critical magnetic field from above. Around the critical magnetic field the one-magnon bands for $q=0,\pi$ can
be well approximated by linearizing the dependence on the field
\begin{align}
\label{eq:1mag_linear}
E_{1m}^\pm&(q=0,\pi)\approx \\
&\frac{1}{2} g_u h_c  -J\epsilon\pm \sqrt{\qty[\frac{J}{2}(1+\epsilon)]^2+(g_sh_c)^2} \nonumber\\
+&\left[ 2g_u \pm h_c \frac{2g_s^2}{\sqrt{J^2\qty(1+\epsilon)^2+4g_s^2 h_c^2}} \right] h. \nonumber
\end{align}
where the critical field $h_c$ is given by
\begin{align}
\label{eq:1mag_hc}
h_c = \frac{J}{ 2\left(g_u^2-g_s^2\right)}& \Big[\epsilon  g_u+ \sqrt{\left(g_u^2-g_s^2\right)(1+\epsilon)^2+4\epsilon^2 g_s^2}  \Big].
\end{align}
For $q=\pi/2$ the one-magnon bands have a linear dependence for any value of the magnetic field,
\begin{align}
\label{eq:1mag_linear2}
E_{1m}^\pm(q=\pi/2)& =-J\epsilon+ \left(g_u \pm  g_s\right) h.
\end{align}
These field-dependent results are superimposed onto the tMPS results presented in Fig.~\ref{fig:summary} above the critical field. The one-magnon
analytical dispersions as a function of the magnetic field strength are denoted by solid blue lines for $q=0, \pi/2, \pi$ and
are in very good agreement with the numerical results obtained when considering the full Hamiltonian.

If one decreases the magnetic field under the critical value, the energy of the lower one-magnon band becomes negative, signaling the
change in nature at the phase transition. However, the upper band still describes well the behavior of the tMPS results for momenta
around $q=0$ and $q=\pi$ with decreasing the magnetic field until $h/J\gtrsim 0.26$, as seen in Fig.~\ref{fig:summary}.
This supports our conclusions of Sec.~\ref{sec:noninteracting} that the high frequency modes have a persistent magnon character
also below the phase transition.

\subsubsection{Two-magnon sector}
\label{sec:2mag}

\begin{figure*}[hbtp]
	\centering
	 \includegraphics[width=0.85\textwidth]{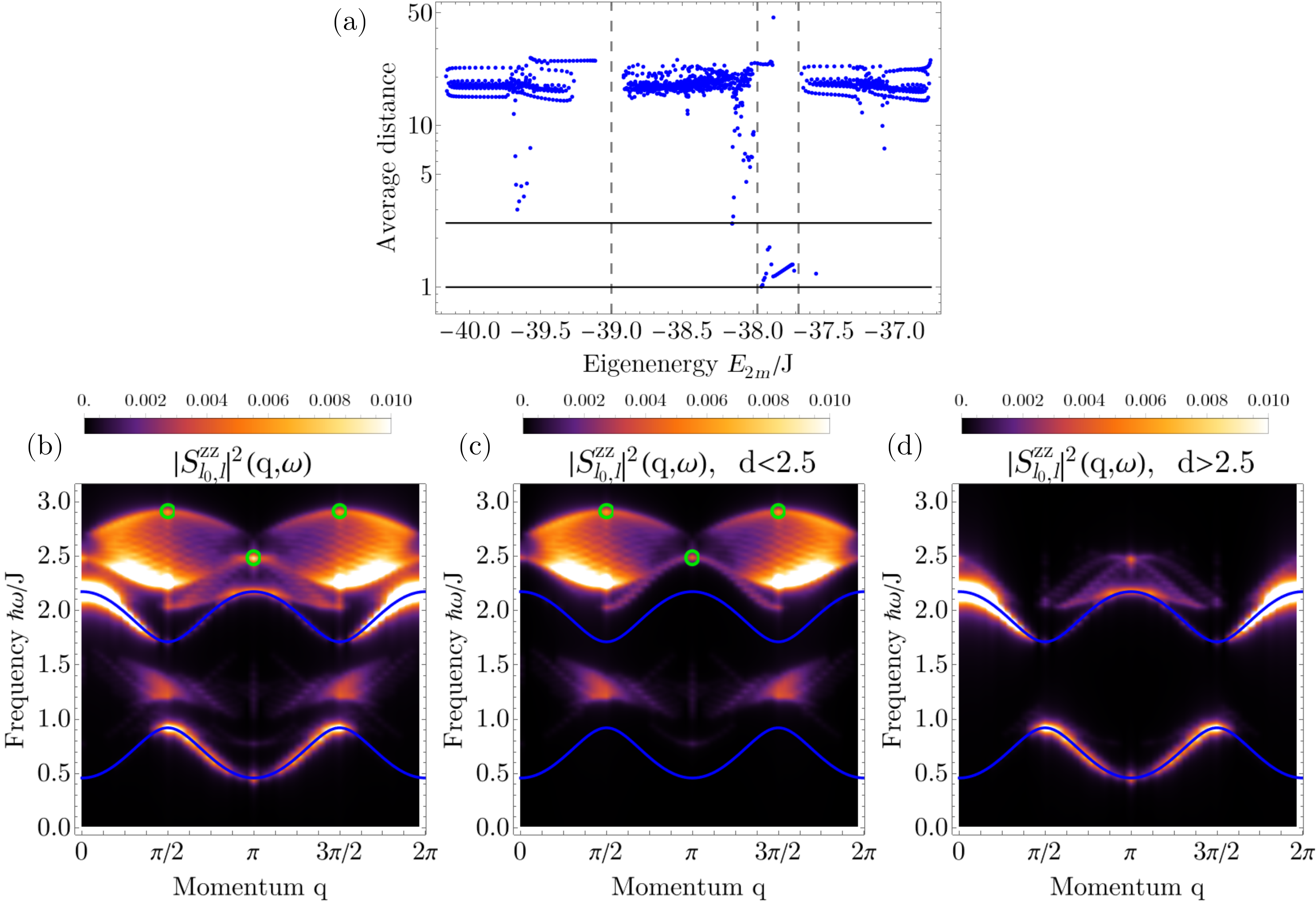} 
	\caption{(a) Average distance between magnons as a function of the eigenenergies for the two-magnon eigenstates, $E_{2m}$. We can identify four distinct bands. The dashed vertical lines mark the boundaries between the two-magnon bands. (b)-(d) Contribution to the dynamical structure factor  due to two-magnon eigenstates $S^{zz}_{l_0,l}(q,\omega)$, in (c)-(d) we have the contribution of eigenstates with an average distance between the two magnons (c) $d < 2.5$, (d) $d > 2.5$ (the $y$-direction is taken within the rotated basis).
The blue line corresponds to the single magnon dispersion, Eq.~(\ref{eq:bands_one_magnon}).
The green circles identify the peaks for $q = \pi/2$ and $q = \pi$. 
Results obtained for the following parameters:$h/J=0.6$ (corresponding to a field $B =56.3~\text{T}$), $L=52$, $g_u=3.06$, $g_s=0.66$, $\epsilon=0.52$, $l_0=26$, $l=27$. }
	\label{fig:2defects}
\end{figure*}

\begin{figure*}[hbtp]
\centering
\includegraphics[width=0.84\textwidth]{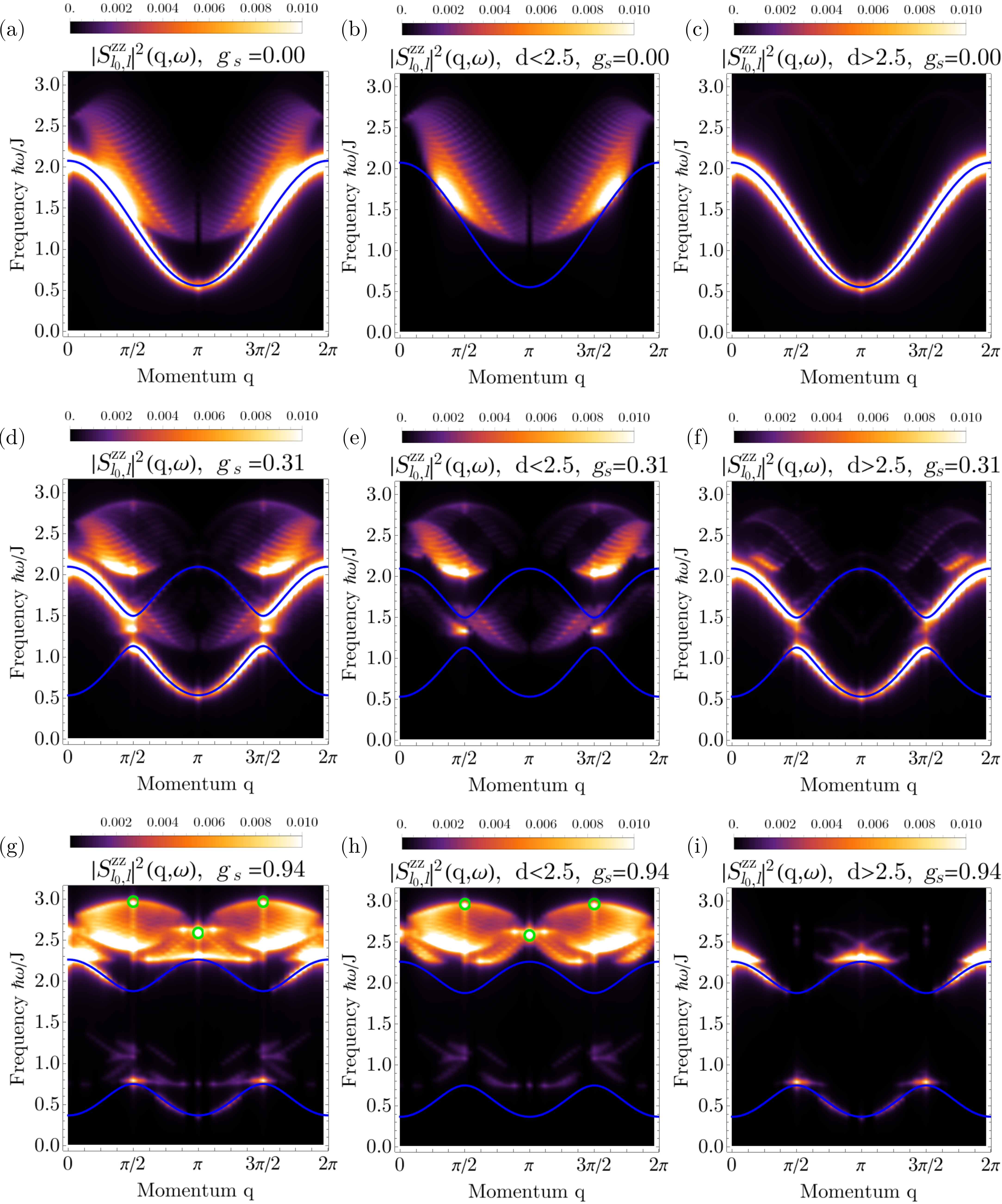} 
\caption{(a), (d), (g) Contribution to the dynamical structure factor  due to two-magnon eigenstates $S^{zz}_{l_0,l}(q,\omega)$ for different values of the staggered magnetic field (the $y$-direction is taken within the rotated basis). In (b), (e), (h) we have the contribution of eigenstates with an average distance between the two magnons $d < 2.5$, and in (c), (f), (i) for a distance $d > 2.5$. The values of the staggering used are (a)-(c) $g_s=0$, (d)-(f) $g_s=0.31$, (g)-(i) $g_s=0.94$.
The blue line corresponds to the single magnon dispersion, Eq.~(\ref{eq:bands_one_magnon}).
Results obtained for the following parameters:$h/J=0.6$ (corresponding to a field $B =56.3~\text{T}$), $L=52$, $g_u=3.06$, $\epsilon=0.52$, $l_0=26$, $l=27$. }
	\label{fig:2defects_staggering}
\end{figure*}

We now turn our attention to the identification of structures appearing at frequencies above the single magnon bands.
Due to the presence of quantum fluctuations for fields below the phase transition threshold, the perturbation acts on states where
flipped spins are already present. Thus, a reasonable candidate for higher modes would include more than one magnons.
The perturbation acting on such a state creates a second magnon via the operator applied at time $t = 0$ within the dynamical
correlation function, Eqs.~(\ref{eq:corr1})-(\ref{eq:corr2}). Since we are mainly interested in contributions from bound states, we start
with an initially localized one-magnon excitation. This situation is described by the dynamical structure factor of the form
\begin{align}
\label{eq:bacovo_corr2def}
S^{\alpha,\beta}_{l_0,l}(q,\omega)&=\sum_{e_{1m}}\sum_{e_{2m}}\braket{l_0}{e_{1m}} \bra{e_{1m}}  S^\alpha(q) \ket{e_{2m}} \times \\
&\times\bra{e_{2m}} S^\beta_l\ket{l_0} \delta(\hbar\omega+E_{1m}-E_{2m}), \nonumber
\end{align}
where $\ket{l_0}$ is the initial state with a localized magnon at site $l_0$,  $\ket{e_{1m}}$ are the one-magnon
eigenstates with the corresponding eigenenergies $E_{1m}$, $\ket{e_{2m}}$ are the two-magnon eigenstates with the corresponding
eigenenergies $E_{2m}$ and $ S^\alpha(q)=\sum_l e^{-iql}S^\alpha_l$. For the calculations presented here,
we considered $\alpha=\beta=y$ in the rotated basis, Eq.~(\ref{eq:bacovo_rot2}) (corresponding to the $z$-direction in the original basis).
We compute the eigenstates $\ket{e_{1m}}$ and $\ket{e_{2m}}$ numerically by performing an exact
diagonalization of the Hamiltonian $\tilde{H}_\text{eff}$, Eqs.~(\ref{eq:Hamiltonian_magnons})-(\ref{eq:Hamiltonian_magnons_fermions}),
in the subspaces of one and two-magnons.

For a two-magnon eigenstate $\ket{e_{2m}}$ we define an average distance between the two magnons
by $d=\bra{e_{2m}}\hat{d}\ket{e_{2m}}=\sum_{l_1<l_2}(l_2-l_1)|\bra{l_1,l_2}\ket{e_{2m}}|^2$,
with $\ket{l_1,l_2}$ representing a state with two magnons at sites $l_1$ and $l_2$.
In Fig.~\ref{fig:2defects}(a), we show the average distance between the two magnons as a function of
the eigenenergies of the Hamiltonian $\tilde{H}_\text{eff}$, Eq.~(\ref{eq:Hamiltonian_magnons}),
in the two-magnon subspace. We observe four distinct bands,
whereas the first, second, and
fourth bands are characterized by an average distance greater than $10$, within the third band, for most eigenstates
the average distance between the two magnons is small, $d<2.5$. Therefore, these eigenstates strongly overlap
with the two-magnon bound states. These bound states result from the presence of magnon-magnon interaction
and appear as a separate band due to the staggering of the magnetic field. 
The sign of magnon-magnon interactions present in the Hamiltonian Eq.~(\ref{eq:Hamiltonian_magnons}) is positive, $J\epsilon>0$, due to the antiferromagnetic character. This implies that the many-body magnon states that we identify are repulsively bound, i.e.~their energy lies above the energy of the unbound states. Repulsively bound state have been observed and discussed previously for atom pairs in optical lattices \cite{WinklerZoller2006,DeuchertCederbaum2012}, where the atoms are well decoupled from the environment.

In Fig.~\ref{fig:2defects}(b), we identify the contributions of the two-magnon eigenstates
to the dynamical structure factor by evaluating Eq.~(\ref{eq:bacovo_corr2def}). For a magnetic field
$h/J=0.6$ (larger than the critical field), we consider a situation where, in the initial state, a magnon is
localized at site $l_0=26$ and a second is one created at $l=27$ under the action of the perturbation. Note that the initial state used
is not an eigenstate and it has finite overlap with both the two-magnon bound states and with the states with delocalized independent magnons.
By comparing Fig.~\ref{fig:2defects}(b), (c)
and (d), we identify the contributions to the dynamical structure factor depending on the average distance between the two magnons.
This is done by splitting the sum over the two-magnon eigenstates. We first identify features which
agree with the single magnon dispersion. This is explained by considering that if the two magnons are
relatively far from each other, their motion is independent while still being affected by the magnetic field staggering.
This is further confirmed by Fig.~\ref{fig:2defects}(d) where we only considered the contributions of the eigenstates
for which the average distance between the magnons is $d>2.5$.

We can also identify many features which are deviating
from the single magnon dispersion, in particular at high frequencies [Fig.~\ref{fig:2defects}(b)(c)]. These are due to the
eigenstates corresponding to two confined magnons as the main contributions to the dynamical structure factor stem from
states for which the two magnons are close to each other. For $d<2.5$ the interaction between the magnons plays an important
role. In fact, we observe in Fig.~\ref{fig:2defects}(c) a high frequency band appearing between
$\hbar\omega/J\approx 2.2$ and $\hbar\omega/J\approx 3$. 
The high-energy nature of the features corresponding to two-magnon bound state, in particular above the single magnon bands, is due to the repulsive nature of the interactions.
We identify two peaks at $q=\pi/2$ and $q=\pi$ in the
dynamical structure factor, marked with green circles in Fig.~\ref{fig:2defects}(c), for which we
monitor the frequency position as a function of the magnetic field. Once again, we can report these peaks
in Fig.~\ref{fig:summary} where the tMPS results for the full Hamiltonian are
presented. One can see that the position of the red solid lines representing these peaks are in very good agreement
with the numerical results obtained when considering the full Hamiltonian.
We therefore conclude that important features of the dynamical structure factor are due to the presence of
two-magnon repulsively bound states. The reduction in weight of these spectral features above the transition is in
agreement with this identification as above the critical field quantum fluctuations become small.

It should be noted that the staggered term of the magnetic field [see, for example, Eq.~(\ref{eq:bacovo_Hamiltonian_reduced_mag})]
plays a very important role in the visibility of the multi-magnon bound states. In order to highlight the role of the staggering, we performed
the same analysis as above (see Fig.~\ref{fig:2defects}) for three different staggering strengths: $g^s = 0$, corresponding to the absence of a staggering field, $g^s = 0.31$,
corresponding to a weaker staggering than for BaCo\textsubscript{2}V\textsubscript{2}O\textsubscript{8} (where the staggering parameter
is $g^s = 0.66$, see Sec.~\ref{sec:numerical_results}), and for $g^s = 0.94$, corresponding to a stronger staggering than in
BaCo\textsubscript{2}V\textsubscript{2}O\textsubscript{8}. The result of this analysis are presented in
Fig.~\ref{fig:2defects_staggering} where, for the three staggering strengths, we show the total dynamical structure factor
due to two-magnon states in the left panels and split, as done above (see Fig.~\ref{fig:2defects}), the contributions in terms of the average distance
between the two-magnons showing the contributions of two-magnon states characterized by an average distance $d<2.5$ in
the central panels and by an average distance $d>2.5$ in the right panels. One first notices in Fig.~\ref{fig:2defects_staggering}(a) and (c)
that, in the absence of staggering, the dynamical structure factor is dominated by a single one-magnon band corresponding to the motion of
a second unbound magnon in a uniform magnetic field. As expected,
in a uniform magnetic field, $g^s = 0$, the pure cosine contribution [see Eq.~(\ref{eq:bands_one_magnon})] is detected. Considering
Fig.~\ref{fig:2defects_staggering}(b), one notices near $q = \pi/2$ that weight associated with two confined magnons is also present even in the
absence of a staggering field. However, this weight occurs at a position where it overlaps with the one-magnon band making its detection difficult.
In the presence of weak staggering ($g^s = 0.31$), one notices considering Fig.~\ref{fig:2defects_staggering}(d) and (f) the
presence of two one-magnon bands corresponding to the motion of a single unbound magnon in a staggered magnetic field background. This excitation
still contributes significantly to the total dynamical structure factor, but contributions due to states corresponding to
two confined magnons can be seen at $q=\pi/2$ both in the gap between the two one-magnon bands and above the upper one.
Finally, for strong staggering strength ($g^s = 0.94$), the contribution of the two-magnon states
to the dynamical structure factor is dominated by states corresponding two confined magnons as can be seen on
Fig.~\ref{fig:2defects_staggering}(g) and (h). For this staggering strength and also for $g^s = 0.66$, corresponding to
BaCo\textsubscript{2}V\textsubscript{2}O\textsubscript{8}, the contributions due to two-magnon bound states are well separated
from the two one-magnon bands both at $q=\pi/2$ and $q=\pi$ making their detection experimentally feasible. Consequently,
the magnetic field staggering plays an essential role as it ensures that the contributions of the two-magnon repulsively bound state
excitations to the structure factor are visible and well separated from the single magnon features.

\subsubsection{Three-magnon sector}
\label{sec:3mag}

\begin{figure*}[hbtp!]
	\centering
	 \includegraphics[width=0.8\textwidth]{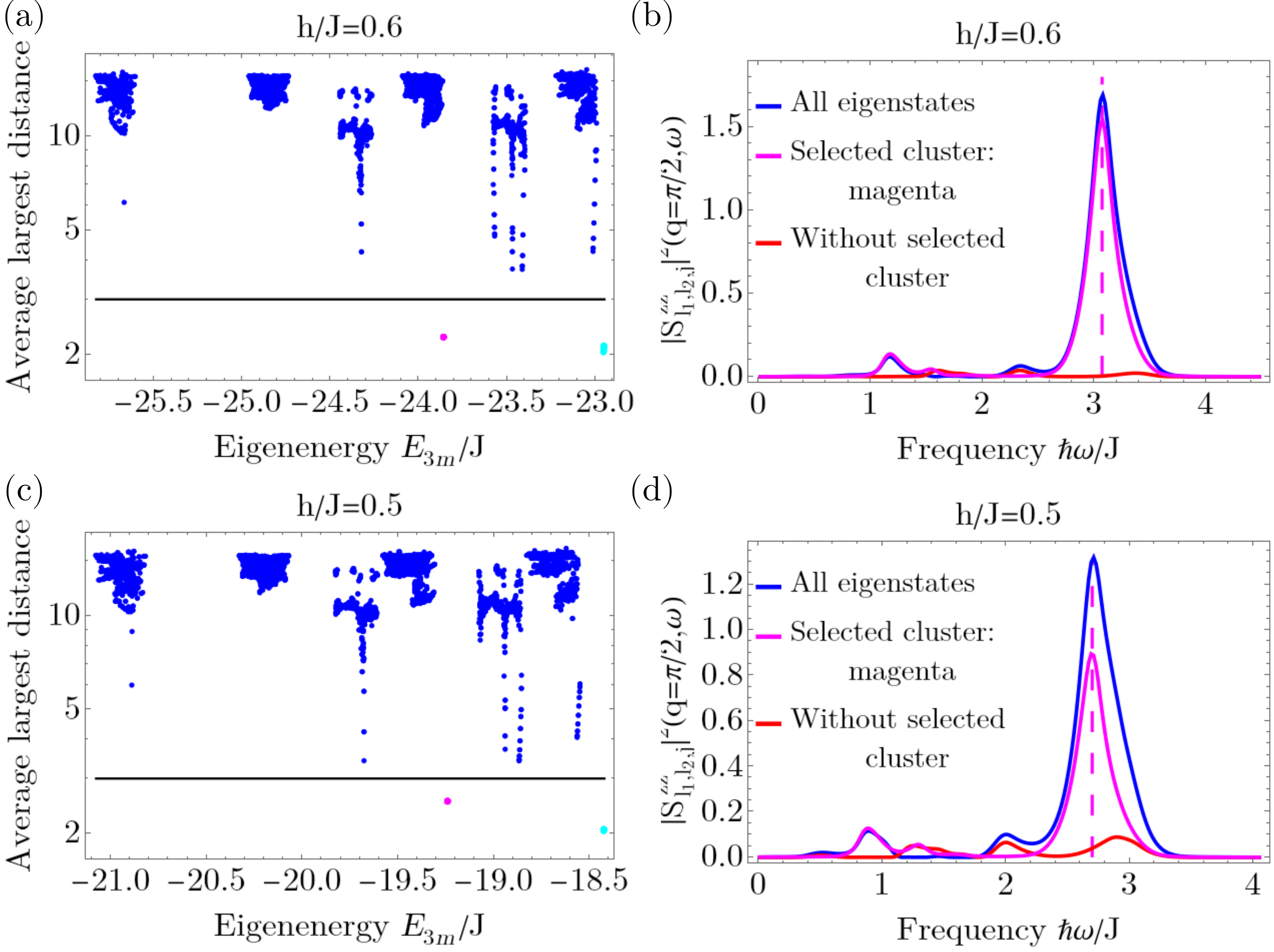} 
	 \caption{(a), (c) Largest average distance between two of the magnons of the three-magnon eigenstates as a function of their corresponding eigenenergies $E_{3m}$.
           The solid black line corresponds to $d = 3$.
	  (b), (d) The dynamical structure factor contribution of the three-magnon eigenstates for $q=\pi/2$ in the $z$-direction (the $y$-direction is taken within the rotated basis).
          We consider the initial state $\ket{l_1=18,l_2=20}$ and $j=19$ and we either take into account all three-magnon eigenstates, the cluster of eigenstates identified by magenta dots in panels (a), (c), or all eigenstates but the selected cluster.
	  We mark the position of the high-frequency peaks which we compare with the numerical tMPS results in Sec.~\ref{comparison}.
	  The cluster identified by cyan dots in panels (a), (c) has been used in the calculations of the dynamical structure factor for the initial state $\ket{l_1=17,l_2=19}$ and $j=18$. Results obtained for the following parameters $h/J=0.6$ (corresponding to a field $B =56.3~\text{T}$) and $h/J=0.5$ (corresponding to a field $B =46.9~\text{T}$), $L=36$ with periodic boundary conditions, $g_u=3.06$, $g_s=0.66$ and $\epsilon=0.52$.}
	\label{fig:3defects}
\end{figure*}

\begin{figure}[hbtp!]
	\centering
	 \includegraphics[width=0.48\textwidth]{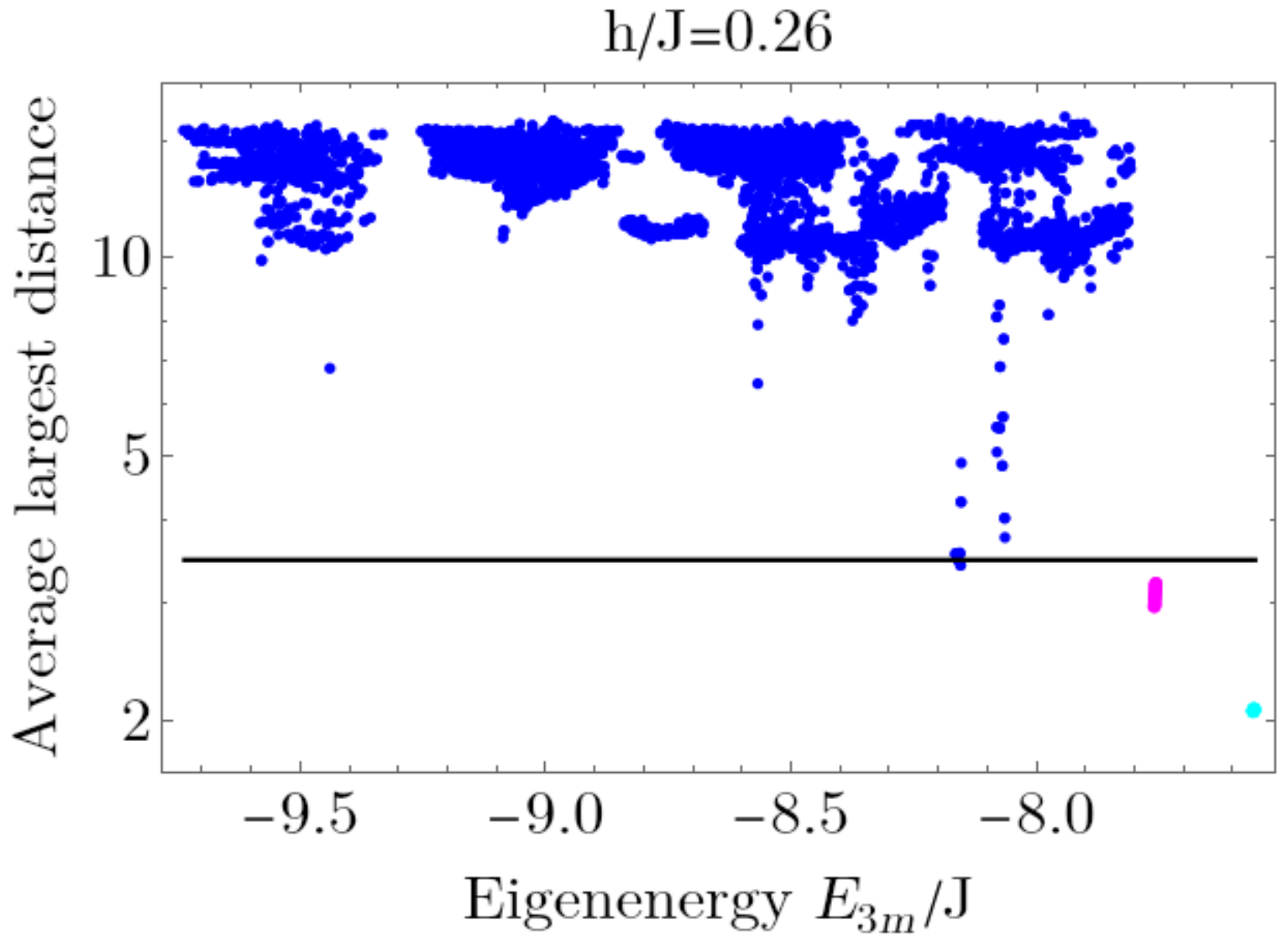} 
	 \caption{Largest average distance between two of the magnons of the three-magnon eigenstates as a function of their corresponding eigenenergies $E_{3m}$.
The solid black line corresponds to $d = 3.5$. Two clusters of eigenstates (identified by magenta dots and cyan dots) for which $d < 3.5$ are found to significantly contribute to the dynamical structure factor. Results obtained for the following parameters $h/J=0.26$ (corresponding to a magnetic field $B =24.4~\text{T}$), $L=36$ with periodic boundary conditions, $g_u=3.06$, $g_s=0.66$ and $\epsilon=0.52$. }
	\label{fig:lowfielddistance}
\end{figure}

Using a similar approach as the one used in the previous section, we now want to understand the dynamical response signature of three-magnon repulsively bound states.
We therefore consider here an initial state with two localized magnons. In this case, the excitations triggered by flipping a spin can be described by the following contribution to the dynamical structure factor
\begin{align}
\label{eq:bacovo_corr3def}
S^{\alpha,\beta}_{l_1,l_2,j}(q,\omega)&=\sum_{e_{2m}}\sum_{e_{3m}}\braket{l_1,l_2}{e_{2m}} \bra{e_{2m}}  S^\alpha(q) \ket{e_{3m}} \nonumber\\
 &\times\bra{e_{3m}} S^\beta_j\ket{l_1,l_2} \delta(\hbar\omega+E_{2m}-E_{3m}), 
\end{align}
where  $\ket{l_1,l_2}$ is the initial state of two localized magnons at sites $l_1$ and $l_2$,  $\ket{e_{2m}}$ are
the two-magnon eigenstates with the corresponding eigenenergies $E_{2m}$, $\ket{e_{3m}}$ are the three-magnon
eigenstates with the corresponding eigenenergies $E_{3m}$. As in Sec.~\ref{sec:2mag}, we consider $\alpha=\beta=y$ in the rotated basis
(corresponding to the $z$-direction in the original basis), and we compute the eigenstates $\ket{e_{2m}}$ and $\ket{e_{3m}}$ numerically by performing
an exact diagonalization of the Hamiltonian $\tilde{H}_\text{eff}$, Eqs.~(\ref{eq:Hamiltonian_magnons})-(\ref{eq:Hamiltonian_magnons_fermions}),
in the subspaces of two and three magnons.

Due to the presence of a staggered magnetic field, three confined magnons can adopt two distinct configurations:
either two of the three magnons are on odd sites, or two are on even sites. In order to take into account both configurations,
we consider two possible initial states $\ket{l_1,l_2}$ in our calculations. We take the two localized magnons either on odd,
or even, sites, with $l_2=l_1+2$, and we create the third magnon at $j=l_1+1$.  We note that by performing the calculations for different initial
states (not shown) we observed that the high frequency modes, which we are interested in, have a larger weight when considering the initial states mentioned above.

In order to identify the three-magnon eigenstates having the largest overlap with configurations corresponding to three
confined magnons, we compute the average distance between magnons for every eigenstate, similar to the analysis performed in Sec.~\ref{sec:2mag}. This distance is defined
as $d=\bra{e_{3m}}\hat{d}\ket{e_{3m}}=\sum_{l_1<l_2<l_3}(l_3-l_1)|\bra{l_1,l_2,l_3}\ket{e_{3m}}|^2$, with $\ket{l_1,l_2, l_3}$
representing a state of three magnons at sites $l_1$, $l_2$ and $l_3$, and corresponds to the largest distance
out of the three possible combinations. In a fully confined state where the three magnons are on consecutive sites, we would obtain $d=2$ as the largest distance. 
However, due to the quantum fluctuations, the three-magnon eigenstates have an overlap also from states with delocalized magnons, resulting in a larger average distance, $d>2$.
Considering a system of length $L=36$ with periodic boundary conditions and for large values of the magnetic field, $h > h_c$, we identify in Fig.~\ref{fig:3defects} (a)(c) two clusters of three-magnon eigenstates (marked by magenta and cyan dots) that strongly overlap with three-magnon bound states as the average largest distance is $d < 3$. In the following, we determine the contribution of these clusters of eigenstates to the dynamical structure factor due to three-magnon excitations.

In Fig.~\ref{fig:3defects} (b)(d), we plot the contribution to the structure factor for $q=\pi/2$, Eq.~(\ref{eq:bacovo_corr3def}),
as a function of frequency for two values of the magnetic field larger than the critical value. In both cases,
we consider the initial state $\ket{l_1=18,l_2=20}$ and take $j=19$. We determine the nature of the modes identified
in Fig.~\ref{fig:3defects} by also computing the dynamical structure factor only including
eigenstates for which the three magnons are most likely confined next to each other. We select the cluster of eigenstates
identified by magenta dots in Fig.~\ref{fig:3defects}(a)(c) as for these eigenstates the largest distance between two
of the three magnons is small, $d<3$. We observe for both values of the magnetic field considered in Fig.~\ref{fig:3defects}(b)(d)
that the main features of the structure factor are dominated by contributions coming from three-magnon bound states. 
The peak positions marked in Fig.~\ref{fig:3defects}(b)(d) are the one reported in Fig.~\ref{fig:summary} to identify the three-magnon features of the dynamical structure factor obtained within tMPS. 
Similarly, we find for the initial state $\ket{l_1=17,l_2=19}$ and with $j=18$ that the cluster of eigenstates identified by the cyan dots in Fig.~\ref{fig:3defects}(a)(c)
plays the most important role and the position of corresponding peaks in the structure factor are shown in Fig.~\ref{fig:summary}.

For lower values of the magnetic field (even for values below the critical field $h_c$), we can also
identify clusters of eigenstates corresponding to states where the three magnons are likely confined together.
For example, for $h/J=0.26$, as shown in Fig.~\ref{fig:lowfielddistance}, we identify three clusters of eigenstates for
which $d < 3.5$ that contribute significantly to the dynamical structure factor. In fact, features of the dynamical structure factor
first identified within the numerically exact tMPS calculations
can still be correctly reproduced using our simplified approach. Thus, it appears that even at lower magnetic fields,
we can identify modes having a significant three-magnon repulsively bound state character at frequencies corresponding to features revealed
via the tMPS numerical simulations (see Fig.~\ref{fig:summary}).

\subsection{Comparison with the numerical results \label{comparison}}

We can compare the approximate analytical results obtained in Sec.~\ref{sec:interacting} with the exact numerical results
described in Sec.~\ref{sec:numerical_results} by tracking the dependence of the different modes when varying the magnetic field,
as shown in Fig.~\ref{fig:summary}. We can observe a very good agreement for all the modes considered, justifying that the approximations
discussed in Sec.~\ref{sec:noninteracting}-\ref{sec:interacting} capture the correct physical behavior.

The one-magnon dispersion describes very well the numerical results above the phase transition, as seen for the
modes $M_\pi^l$, $M_0^u$, $M_{\pi/2}^l$ and $M_{\pi/2}^u$ in Fig.~\ref{fig:spectra} and Fig.~\ref{fig:summary} (analytical results depicted with blue curves). 
The lower modes  $M_\pi^l$ and $M_{\pi/2}^l$ change their character below the transition and cannot be described anymore by the single magnon excitation.
In contrast, the higher excited mode $M_0^u$ is captured by the one-magnon curve also below the phase transition [Fig.~\ref{fig:summary}(a),(d)].
This is in agreement with our findings of section~\ref{sec:analytics} that the single magnon character of the higher energy mode survives
well below the transition, whereas the low energy excitations change their character.

For the case of the high-energy modes, we observed in Fig.~\ref{fig:summary} a good agreement between the
modes $D_{\pi/2}$ and $D_{\pi}$, and the red curves corresponding to the analytical results for the two-magnon bound states.
 Above the transition, the weight obtained within the numerical calculations and associated with the
  two-magnon bound states is low as the ground state is the initial state. In this state, only a few magnons
  are readily available due to zero-point fluctuations and can then form
  a two-magnon excitation with the magnon created by the probing field (the operator applied on the initial state within the
  structure factor formalism).
At finite (but low) temperature, we would expect that the weight associated with these modes would increase and that two-magnon states become
more visible also above the transition. Below the transition, more magnon states are already present which leads to considerably
stronger weight for the two-magnon states.
 Additionally, at $q=\pi/2$, a light feature can be seen in the structure factor calculated numerically via tMPS. This feature is
  nicely following the $T_{\pi/2}$ mode indicated by the magenta curve obtained analytically when considering three-magnon bound states.
  As such, we attribute the origin of this feature to three-magnon bound states. We have thus identified and characterized most of the modes
  present in the spectral function of the full model by the considering simpler approximate models.

\section{Conclusion \label{sec:conclusion}}

In summary, we investigated the excitation spectrum of a spin-1/2 XXZ chain with antiferromagnetic Ising anisotropy in a magnetic
field with a strong transverse $x$-direction component with both uniform and staggered contributions and a weak four-fold periodic $z$-direction longitudinal component.
Motivated by the experimental results on the spin-1/2 chain antiferromagnet BaCo\textsubscript{2}V\textsubscript{2}O\textsubscript{8} compound \cite{WangKollath2024},
we considered the dominant contribution to the site-dependent magnetic field to be the uniform field in the $x$-direction with an additional staggered component.
We used the time-dependent matrix product state method to compute the spin dynamical structure factor in the regime of strong magnetic fields across
the phase transition from the antiferromagnetic state in $z$-direction to the field polarized state in the $x$-direction.
The critical point of the phase transition can be identified in the structure factor by the closing of the gap at the transition.

The numerically exact tMPS results are complemented by analytical calculations which allow us to determine the nature of the excitations.
Above the phase transition threshold at zero-temperature the dynamics of the model is dominated by single magnons, i.e.~spin flips on top of
a polarized background, moving in a potential with a staggered amplitude given by the external magnetic field.
Furthermore, we show that for strong magnetic fields below the transition threshold, the high-energy excitations can still be
understood in terms of magnons in the direction of the applied external field.
We identify many-body repulsively bound magnon state structures within the dynamical structure factor.
These are states in which two, or three, magnons are confined together and they arise due to the interplay between
the many-body repulsive interactions and the magnetic field. The two- and three-magnon bound states
can be distinguished as separate modes in the excitation spectrum due to the presence of the staggered component
in the transverse magnetic field. 
In an experimental solid state realization of this model, the BaCo\textsubscript{2}V\textsubscript{2}O\textsubscript{8} compound,
  the repulsively bound states might have a metastable character. Previously, such repulsively bound states have been identified in
  ultracold quantum gases, systems which are well isolated from their environments \cite{WinklerZoller2006,DeuchertCederbaum2012}.
  Thus, it is fascinating that the repulsively bound states are clearly detected in a real material \cite{WangKollath2024}, where intricate
  couplings between the different degrees of freedom are likely present. It is an open question how stable these repulsively bound
  states are in BaCo\textsubscript{2}V\textsubscript{2}O\textsubscript{8} and other realistic compounds. Besides the mentioned modes,
the nature of low-frequency excitations below the phase transition, which can no longer be described in terms of magnons, remains
to be elucidated in future work.

The parameters considered in this work corresponds to the regime for which BaCo\textsubscript{2}V\textsubscript{2}O\textsubscript{8}
can be described as a quasi-one-dimensional system with weak interchain couplings \cite{WangLoidl2018b}. Thus, our
result can be directly used to identify the nature of the excitation modes measured via terahertz spectroscopy in Ref.~\cite{WangKollath2024},
where a very good agreement between the numerically exact results and the experimental data is found.

\section*{Acknowledgments}

We thank M.~Garst, T.~Giamarchi, S.~Wolff, J.~Wu, and H.~Zou for helpful discussions.
We acknowledge support by the Deutsche Forschungsgemeinschaft (DFG) via TRR80 (subproject F5), CRC1238 (subprojects A02, B01, B05 and C05), TRR 185 (subproject B3), SFB 1143 (subproject C01), by the European Research Council (ERC) under the Horizon 2020 research and innovation programme, grant agreement No. 648166 (Phonton) and No. 950560 (DynaQuanta), by the Natural Sciences and Engineering Research Council of Canada (NSERC) [funding references No. RGPIN-2021-04338 and No. DGECR-2021-00359], and by the Swiss National Science Foundation under Division II grant 200020-188687.

\section*{APPENDIX}

\setcounter{section}{0}
\renewcommand{\thesection}{\Alph{section}}
\renewcommand{\theequation}{A.\arabic{equation}}
\setcounter{equation}{0}

\subsection{BaCo\textsubscript{2}V\textsubscript{2}O\textsubscript{8} $g$-factors\label{app:gfactors}}
The space dependence of the magnetic field given in Eq.~(\ref{eq:Hamiltonian2}) has been derived based on the anisotropic Landé $g$-factors of the BaCo\textsubscript{2}V\textsubscript{2}O\textsubscript{8} compound (see Ref.~\cite{KimuraWatanabe2013} and  Supplemental Material of Ref.~\cite{WangKollath2024}).
The non-zero entries relevant for the determination of the effective magnetic field are given by
\begin{align}
\label{eq:bacovo_0direction}
g^{xx}_j&=\qty(g_1\cos^2\theta+g_2\sin^2\theta)\cos^2\qty(\frac{\pi}{2}(j-1)) \\
&\qquad+g_3\sin^2\qty(\frac{\pi}{2}(j-1)), \nonumber\\
g^{xz}_j &=(g_2 - g_1)\cos\theta\sin\theta\cos\qty(\frac{\pi}{2}(j-1)), \nonumber 
\end{align}
where $\theta=5^\circ$ is the tilt angle and $g_1$, $g_2$, and $g_3$ are the values of the $g$-tensor along the magnetic principle axes.

\end{document}